\documentclass[twocolumn, floatfix]{aastex701}

\usepackage[T1]{fontenc}
\usepackage[utf8]{inputenc}
\usepackage{amsmath,amssymb}
\usepackage{graphicx}
\usepackage{booktabs}
\usepackage{array}
\usepackage{bm}
\usepackage{placeins}
\usepackage{xcolor}
\usepackage{hyperref}
\usepackage{tabularx}

\newcommand{\Mstar}{\ensuremath{M_{\star}}}
\newcommand{\Lz}{\ensuremath{L_{z}}}
\newcommand{\Jang}{\ensuremath{J}}
\newcommand{\Mzgas}{\ensuremath{M_{Z,\mathrm{gas}}}}
\newcommand{\sSFR}{\ensuremath{\mathrm{sSFR}}}

\newcommand{\logsSFR}{\ensuremath{\log(\sSFR)}}

\newcommand{\ssfrquenchthr}{-2.0}

\begin{document}

\title[GNN Surrogate for Semi-analytic Galaxy Evolution]{A graph-based Neural Network surrogate model for accelerating semi-analytical model of galaxy formation and evolution}

\correspondingauthor{Zhongxu Zhai}
\email{zhongxuzhai@sjtu.edu.cn}

\correspondingauthor{Xiaohu Yang}
\email{xyang@sjtu.edu.cn}

\author{Xuejie Li}
\email{lixuejie@sjtu.edu.cn}
\affiliation{State Key Laboratory of Dark Matter Physics, Tsung-Dao Lee Institute \& School of Physics and Astronomy, Shanghai Jiao Tong University, Shanghai 200240, China}
\affiliation{Shanghai Key Laboratory for Particle Physics and Cosmology,  and Key Laboratory for Particle Physics, Astrophysics and Cosmology, Ministry of Education, Shanghai Jiao Tong University, Shanghai 200240, China}

\author{Zhongxu Zhai}
\email{zhongxuzhai@sjtu.edu.cn}
\affiliation{State Key Laboratory of Dark Matter Physics, Tsung-Dao Lee Institute \& School of Physics and Astronomy, Shanghai Jiao Tong University, Shanghai 200240, China}
\affiliation{Shanghai Key Laboratory for Particle Physics and Cosmology,  and Key Laboratory for Particle Physics, Astrophysics and Cosmology, Ministry of Education, Shanghai Jiao Tong University, Shanghai 200240, China}
\affiliation{Waterloo Center for Astrophysics, University of Waterloo, Waterloo, ON N2L 3G1, Canada}
\affiliation{Department of Physics and Astronomy, University of Waterloo, Waterloo, ON N2L 3G1, Canada}

\author{Xiaohu Yang}
\email{xyang@sjtu.edu.cn}
\affiliation{State Key Laboratory of Dark Matter Physics, Tsung-Dao Lee Institute \& School of Physics and Astronomy, Shanghai Jiao Tong University, Shanghai 200240, China}
\affiliation{Shanghai Key Laboratory for Particle Physics and Cosmology,  and Key Laboratory for Particle Physics, Astrophysics and Cosmology, Ministry of Education, Shanghai Jiao Tong University, Shanghai 200240, China}

\author{Andrew Benson}
\email{abenson@carnegiescience.edu}
\affiliation{Carnegie Observatories, 813 Santa Barbara Street, Pasadena, CA 91101, USA}

\author{Yun Wang}
\email{wang@ipac.caltech.edu}
\affiliation{IPAC, California Institute of Technology, Mail Code 314-6, 1200 E. California Blvd., Pasadena, CA 91125, USA}

\begin{abstract}
Understanding how galaxy populations emerge and evolve from the growth of dark matter structure is a central challenge in galaxy formation theory. Semi-analytic models (SAMs) provide an efficient framework {to address} this problem, but exploring large ensembles of merger trees across broad parameter spaces remains computationally demanding. We develop a conditional graph neural network surrogate model that combines merger tree information with SAM parameters to predict galaxy properties across cosmic time. Using merger trees of dark matter halos from the Uchuu simulation and the \textsc{Galacticus} SAM, the model predicts stellar mass, luminosity, angular momentum, gas metal mass, and specific star formation rate across {the} wide redshift range {of $0\leq z\la 5$}. For instance, the model can predict stellar mass at {$0\leq z \la 3$} with a scatter of 
{0.19-0.28} dex and 
{coefficient of determination} {$R^2$ of 0.946-0.973 ($R^2$ close to 1 indicates prediction closely matching the truth)}. The results show that a single graph based model can reproduce these galaxy properties with good accuracy over multiple SAM realizations, merger trees and redshifts. This catalog-level model provides a practical route for accelerating SAM based studies of galaxy formation to enable a more detailed investigation of the model parameter space. The inference code, trained models, {{and example data products are publicly available at \url{https://doi.org/10.5281/zenodo.21367843}}}.

\end{abstract}

\keywords{
galaxies: formation -- galaxies: evolution -- methods: data analysis -- methods: numerical
}

\section{Introduction}

In our current understanding of galaxy formation and evolution, galaxy properties are shaped by the coupled effects of halo assembly and baryonic evolution across cosmic time. Gas accretion, cooling, star formation, feedback, mergers, and chemical enrichment all contribute to this connection, so a useful forward model must link halo growth to galaxy populations in a physically meaningful way \citep{benson2010galaxy, somerville2015physical}. The current model of galaxy formation is built on this principle and has achieved substantial success in reproducing many observed trends \citep{vogelsberger2014illustris, schaye2015eagle, pillepich2018first}, but it remains far from complete in terms of theoretical description \citep{benson2010galaxy, somerville2015physical} and detailed comparisons with observational data \citep{henriques2009mcmc, lu2011bayesian, vernon2014galaxy, robertson2026accelerated}.

More broadly, the relation between galaxies and dark matter halos has long been studied with a variety of empirical and statistical models, including halo occupation distributions, conditional luminosity functions, abundance matching, and related empirical galaxy--halo connection frameworks; see \citet{Wechsler_Tinker_2018} for a review, and \citet{zheng2005hod, yang2003clf, moster2013shmr, rodriguez2017shmr} for representative examples. Among physically motivated forward models, two of the main approaches are hydrodynamical simulations and semi-analytic models (SAMs). Hydrodynamical simulations such as Illustris \citep{vogelsberger2014illustris}, EAGLE \citep{schaye2015eagle}, and IllustrisTNG \citep{pillepich2018first} evolve a tremendous number of particles of dark matter, gas, and stars self-consistently and have substantially improved comparisons between theory and observation.
 SAMs instead evolve galaxies in the background of dark matter halos and their merger trees using parameterized prescriptions for cooling, star formation, feedback, enrichment, and satellite evolution, and therefore offer greater flexibility for parameter studies and mock catalog construction \citep{white1991galaxy, cole2000hierarchical, benson2012galacticus, somerville2015physical}.

Although SAMs are flexible, many applications still require repeated model evaluations over broad parameter spaces for calibration and inference \citep{henriques2009mcmc, lu2011bayesian, vernon2014galaxy, Gomez_2025, robertson2026accelerated}. {A common used approach for calibrating SAM parameters relied on Monte Carlo merger trees generated from extended Press–Schechter-type models, instead of using the full merger trees obtained from cosmological simulations. This is because such Monte Carlo trees are inexpensive to produce and enable a fast exploration of the parameter space \citep{parkinson2008trees, lu2011bayesian, robertson2026accelerated}.} This approach has been widely used to calibrate SAM parameters by comparing predicted galaxy abundance statistics with observations, for example, through the stellar mass function and luminosity function \citep{henriques2009mcmc, lu2011bayesian, vernon2014galaxy}. However, such merger trees are not taken directly from simulations; they are less accurate for applications that require simulation-based assembly histories of dark matter halos or information about their spatial clustering \citep{zhai2025clustering}. {For such problems, SAMs constructed from a small set of merger trees derived from cosmological simulations are often the more suitable option. Nevertheless, thoroughly probing the broad SAM parameter space via repeated forward modeling remains computationally costly.} This is the regime in which surrogate models are particularly well motivated.

Machine learning has become an increasingly important tool for accelerating expensive forward models in galaxy formation and cosmology. This includes applications to simulation emulation, fast nonlinear prediction, and data-driven approximations to complex physical evolution \citep{he2019learning, de2020fast, dai2021learning}. Early studies showed that supervised {machine learning} models could recover galaxy properties from dark matter halo information and from semi-analytic catalogs with reasonable accuracy \citep{xu2013first, kamdar2016machine}. Machine learning has also been used to model the galaxy--halo connection in a more flexible way, beyond prescriptions based solely on halo mass, by incorporating secondary halo properties such as concentration, assembly history, and environment \citep{delgado2022mlghc}. Subsequent works used machine learning to populate halos with galaxies, construct fast emulators, and accelerate downstream inference tasks \citep{agarwal2018painting, jo2019mssm, elliott2021efficient, lovell2022machine, he2019learning, de2020fast, ni2021ai}. These developments have established that learned models can capture nontrivial mappings between dark matter structure, baryonic physics, and observable galaxy properties, and can therefore serve as practical surrogates in settings where repeated model evaluations are required. Recent CAMELS-based work has further highlighted the breadth of galaxy formation parameter space and the nonlinear coupling among astrophysical parameters, cosmology, and galaxy populations \citep{ni2023camels}. Together, these developments show that learned surrogates can play a practical role in galaxy modeling pipelines built on large samples.

For SAM emulation, the way in which the input is represented can affect the information that the model can use \citep{hamilton2017inductive, jespersen2022mangrove, chuang2024leaving}. Galaxy properties at a given redshift depend not only on the instantaneous halo state but also on the hierarchical assembly history of the host halo, encoded by the merger tree \citep{behroozi2019universemachine}. Using only a tabulated set of halo properties or summary statistics may therefore lose information that is relevant to the prediction task. Since merger trees are non-Euclidean objects with explicit progenitor--descendant relations, graph neural networks (GNNs) can be a natural candidate for learning on such structured inputs. GNNs operate by passing and aggregating information along graph edges, allowing the representation of each node to depend on its local neighborhood, through repeated message passing, and on the broader graph structure \citep{hamilton2017inductive}. This makes them well-suited to merger trees, where the relevant information is distributed across the full assembly history rather than concentrated on a small set of summary variables. Recent work has shown that GNNs can learn directly from merger trees and predict galaxy properties from halo assembly histories \citep{jespersen2022mangrove, chuang2024leaving}. More broadly, related graph-based generative models also suggest that graph representations are useful for galaxy population modeling beyond standard regression settings \citep{jagvaral2022galaxies, jagvaral2022ia, desanti2025galaxy}.

Existing graph-based studies of merger tree learning have so far focused on more specialized settings, typically within a fixed dataset or simulation setup, rather than across multiple SAM parameter choices and catalogs \citep{jespersen2022mangrove, chuang2024leaving}. An ideal surrogate intended for SAM-based parameter studies should instead ingest both the merger tree and the SAM parameter vector, then return several galaxy properties over multiple redshift outputs. Such a model should generalize simultaneously over halo assembly histories, across catalogs induced by different SAM parameters, and correlations among multiple galaxy properties across cosmic time. This setting, with multiple SAMs, multiple redshifts, and multiple properties, is the central problem considered in this paper.

In this work, we develop a conditional GNN surrogate for the \textsc{Galacticus} SAM that takes halo merger trees together with SAM parameters and predicts stellar mass, luminosity, angular momentum, specific star formation rate, and gas metal mass at multiple redshifts. The model is designed to work across multiple \textsc{Galacticus} catalogs rather than within a single realization, and to treat smooth targets and targets with more complex conditional distributions within the same framework. Our aim is not only to reproduce these quantities accurately, but also to compare how predictive performance differs across targets with different dependence on merger history, SAM parameter variation, and state structure.
{This is part of the work effort in building a robust theoretical framework to interpret the great wealth of galaxy properties data from the Euclid mission \citep{Euclid} and the Nancy Grace Roman Space Telescope \citep{Spergel2015,Wang2022}.}

The remainder of the paper is organized as follows. Section~\ref{sec:data} describes the simulation products, the \textsc{Galacticus} targets, and the data setup. Section~\ref{sec:methodology} presents the graph construction and model architecture, and Section~\ref{sec:training} describes the training objective. Section~\ref{sec:results} reports the main predictive results, while Section~\ref{sec:discussion} discusses their interpretation and limitations. {We summarize and conclude in} Section~\ref{sec:conclusion}.

\section{Data}\label{sec:data}
\subsection{Halo Merger Trees}
Our inputs are halo merger trees extracted from the UchuuMicro $N$-body simulation of the \textsc{Uchuu} suite \citep{ishiyama2021uchuu}. The simulation volume used here has a box size of $100\,h^{-1}\,\mathrm{Mpc}$. Dark matter halos are identified with the \textsc{ROCKSTAR} halo finder \citep{behroozi2013rockstar}, and merger trees are constructed with \textsc{Consistent Trees} \citep{behroozi2013consistent}. In order to build the GNN model, we represent the resulting halo population as a graph in which nodes correspond to halos at individual simulation outputs. The graph includes both progenitor--descendant links, which encode assembly history across time, and host--satellite links, which encode the relation between a subhalo and its host halo at fixed output time. In the full graph used in this work, the halo population comprises approximately $4.7\times10^{6}$ nodes, constructed by tracing halos across multiple snapshot times (defined in Section~\ref{sec:methodology}). Each halo node is associated with a 14-dimensional input vector,
\begin{equation}
    \mathbf{x}_{\rm halo} =
    \left(
    \widetilde{J}_x,
    \widetilde{J}_y,
    \widetilde{J}_z,
    \log M_{\rm halo},
    \mathbf{r},
    \mathbf{v},
    r_{\rm s},
    \lambda,
    a
    \right),
    \label{eq:halo_features}
    \end{equation}
where \(\widetilde{J}_x\), \(\widetilde{J}_y\) and \(\widetilde{J}_z\) denote signed logarithmic transforms of the three halo angular momentum components, \(M_{\rm halo}\) is the halo mass, \(\mathbf{r}\) and \(\mathbf{v}\) are the three dimensional position and velocity vectors respectively, \(r_{\rm s}\) is a scale radius, \(\lambda\) is the spin parameter, and \(a\) is the 
{cosmic scale factor.}
Specifically, for each angular momentum component \(J_i\) we define
\begin{equation}
    \widetilde{J}_i =
    \begin{cases}
    \log_{10} J_i, & J_i > 0,\\
    0, & J_i = 0,\\
    -\log_{10}|J_i|, & J_i < 0.
    \end{cases}
    \label{eq:halo_ang}
\end{equation}
Indeed, in the halo graph used for training and evaluation, we do not encounter components with {\({|J_i|<1|}\)}, so this transform does not introduce an ambiguity between small positive values and intrinsically negative components.

\subsection{\textsc{Galacticus} Catalogs}
\begin{table}
  \centering
  \caption{The parameters of the \textsc{Galacticus} SAM varied in this work, together with their physical meaning and the ranges adopted in the mock suite.}
  \label{tab:sam_parameters}
  \small
  \setlength{\tabcolsep}{3pt}
  \begin{tabular}{lll}
    \toprule
    Parameter & Physical meaning & Range \\
    \midrule
    $\nu_{\mathrm{SF},0}$ & \parbox[t]{0.52\columnwidth}{Normalization of the quiescent star formation law} & $[10^{-10},\,10^{-8}]$ \\
    $\epsilon_\star$ & \parbox[t]{0.52\columnwidth}{Efficiency parameter for star formation in the spheroid} & $[0.01,\,1]$ \\
    $\alpha_\star$ & \parbox[t]{0.52\columnwidth}{Exponent parameter for star formation in the spheroid} & $[-3,\,3]$ \\
    $\beta_{\max,\mathrm{d}}$ & \parbox[t]{0.52\columnwidth}{Maximum of the mass loading factor in the disk} & $[100,\,1000]$ \\
    $V_{\mathrm{outflow},\mathrm{d}}$ & \parbox[t]{0.52\columnwidth}{Velocity of the outflow in the disk} & $[10,\,300]$ \\
    $\alpha_{\mathrm{outflow},\mathrm{d}}$ & \parbox[t]{0.52\columnwidth}{Power law index of the disk outflow prescription} & $[1,\,4]$ \\
    $\beta_{\max,\mathrm{s}}$ & \parbox[t]{0.52\columnwidth}{Maximum of the mass loading factor in the spheroid} & $[100,\,1000]$ \\
    $V_{\mathrm{outflow},\mathrm{s}}$ & \parbox[t]{0.52\columnwidth}{Velocity of the outflow in the spheroid} & $[10,\,300]$ \\
    $\alpha_{\mathrm{outflow},\mathrm{s}}$ & \parbox[t]{0.52\columnwidth}{Power law index of the spheroid outflow prescription} & $[1,\,4]$ \\
    $f_{\mathrm{core}}$ & \parbox[t]{0.52\columnwidth}{Core radius of the hot halo in units of the virial radius of the host dark matter halo} & $[0.01,\,0.8]$ \\
    $f$ & \parbox[t]{0.52\columnwidth}{Age factor for the computation of the time available for cooling} & $[0,\,1]$ \\
    $c_{\mathrm{multi}}$ & \parbox[t]{0.52\columnwidth}{Multiplicative factor controlling the hot gas density profile} & $[0.1,\,5]$ \\
    $\gamma$ & \parbox[t]{0.52\columnwidth}{$\gamma$ in the parametrized model for outflow reincorporation} & $[0.05,\,20]$ \\
    $\delta_1$ & \parbox[t]{0.52\columnwidth}{$\delta_1$ in the parametrized model for outflow reincorporation} & $[0.3,\,5]$ \\
    $\delta_2$ & \parbox[t]{0.52\columnwidth}{$\delta_2$ in the parametrized model for outflow reincorporation} & $[0.3,\,5]$ \\
    $\eta_{\mathrm{radio}}$ & \parbox[t]{0.52\columnwidth}{Efficiency parameter for AGN feedback in the radio mode} & $[0.01,\,3]$ \\
    $\gamma_f$ & \parbox[t]{0.52\columnwidth}{Amplitude of the halo velocity field relative to the raw simulation output} & $[0.5,\,1.5]$ \\
    \bottomrule
  \end{tabular}
\end{table}

Galaxy properties are generated with the \textsc{Galacticus} semi-analytic model \citep{benson2012galacticus}. Given a halo merger tree and a set of SAM parameters, \textsc{Galacticus} evolves the baryonic components of the system using parameterized prescriptions for gas accretion and cooling, star formation, stellar and AGN feedback, chemical enrichment, and angular momentum evolution. In the present work, the \textsc{Galacticus} outputs define the target quantities to be predicted from the halo merger trees.

A key feature of our data set is that it contains many \textsc{Galacticus} realizations generated from different SAM parameter choices. This allows us to study not only variation from one halo assembly history to another, but also the change in predicted galaxy properties induced by different semi-analytic prescriptions. Throughout the paper, we refer to these realizations as separate \textsc{Galacticus} catalogs.

Each catalog is labeled by a 17-dimensional parameter vector. The varied parameters control the normalization of star formation, the strength and scaling of outflows driven by stellar feedback, the reincorporation of expelled gas, the structure and cooling of the hot halo, and the efficiency of radio mode AGN feedback. The individual parameters and the ranges adopted in this work are summarized in Table~\ref{tab:sam_parameters}; their physical interpretation follows the \textsc{Galacticus} parameterization described by \citet{zhai2025clustering}, and we refer the readers to this paper for more details.

\subsection{Target Quantities and Redshift Sampling}\label{sec:targets}
\begin{figure*}
  \centering
  \includegraphics[width=0.72\textwidth]{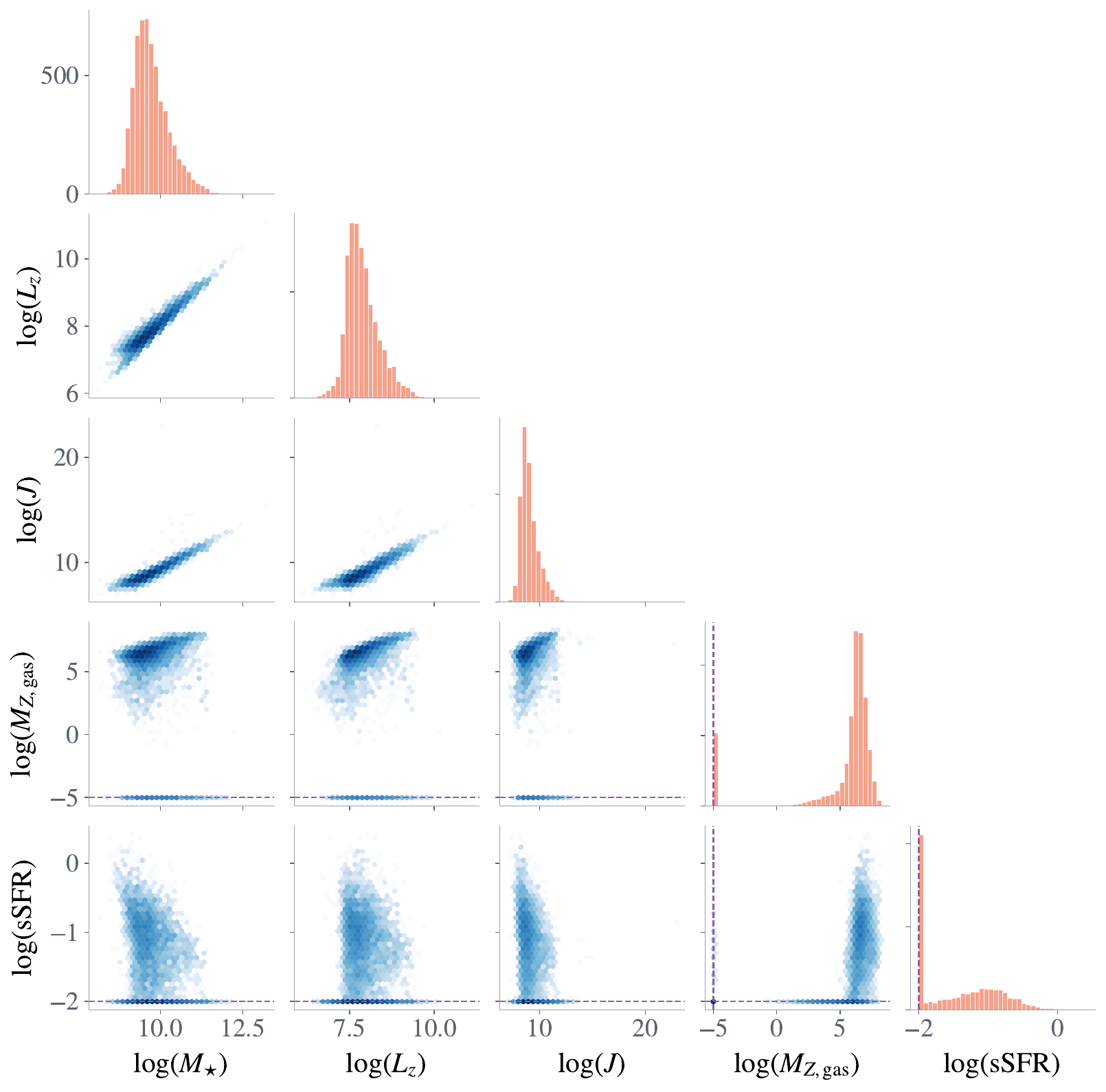}
  \caption{Lower triangular pair plot for the five target properties at $z_1=0$ in a representative single \textsc{Galacticus} catalog drawn from the halo mass selected sample used in the main analysis, illustrating the joint target distribution.}
  \label{fig:fig07}
\end{figure*}
For each target system, we extract galaxy properties at nine output redshifts. Throughout the paper, we index the ordered snapshot sequence by $t=1,\dots,9$, and denote the corresponding redshifts by
\[
\begin{aligned}
\{z_t\}_{t=1}^{9} = \{&
0.00,\,0.25,\,0.49,\,0.70,\,1.03,\\
&
1.54,\,2.03,\,2.95,\,5.15
\}.
\end{aligned}
\]
This notation is used consistently in the text, figures, and tables for each redshift.

At each redshift, we predict the following five galaxy properties:
\begin{itemize}
    \item \textbf{Stellar mass} ($\log_{10}(M_\star/M_\odot)$), the total stellar mass of the galaxy.
    \item \textbf{SDSS $z$-band luminosity} ($\log_{10}(\Lz/\mathrm{W\,Hz^{-1}})$), the total  luminosity output by \textsc{Galacticus} in the SDSS $z$ filter in the observer frame.
    \item \textbf{Angular momentum} ($\log_{10}(\Jang/\mathrm{kg\,m^2\,s^{-1}})$), using the same definition and units as the \textsc{Galacticus} output.
    \item \textbf{Specific star formation rate} ($\log_{10}(\sSFR/\mathrm{Gyr}^{-1})$), defined as $\sSFR=\mathrm{SFR}/\Mstar$.
    \item \textbf{Gas metal mass} ($\log_{10}(M_{\mathrm{Z,gas}}/M_\odot)$), the total metal mass in the gas component.
\end{itemize}
These targets span several aspects of galaxy evolution: integrated stellar growth through $\Mstar$, photometric output through $\Lz$, dynamical information through $\Jang$, recent star formation activity through \sSFR, and chemical enrichment through $\Mzgas$. Figure~\ref{fig:fig07} shows a representative pair plot of these five targets at $z_1=0$, illustrating both their marginal distributions and their correlations within a single \textsc{Galacticus} catalog.

\subsection{Classification Labels}\label{sec：classification_label}
We note that some targets contain a substantial fraction of exact zeros for some parameters, for instance, the star formation rate in the raw \textsc{Galacticus} output. Before applying logarithmic conversion to strictly non-negative quantities, we map these zeros to a fixed floor value of $-5$. This treatment is primarily a numerical convention that allows the targets to remain in log space while preserving the distinction between exact zeros and strictly positive values.
We define two binary labels that are used in the classification summaries reported later in the paper: a quenched label based on \sSFR{} and a label for the floor state in gas metal mass.

\paragraph*{Quenched label:}\ 
We use the term \emph{quenched} for systems with low specific star formation rate. Operationally, a galaxy is labeled as quenched at redshift $z_t$ if
\[
\log_{10}\!\left[\mathrm{sSFR}(z_t)/\mathrm{Gyr}^{-1}\right] \le \ssfrquenchthr.
\]
This classification is used because the \sSFR\ distribution contains a pronounced low-\sSFR\ population, especially at later times \citep{ilbert2013mass, houston2023using}. In addition, our preliminary inspection of the catalog ensemble shows that the quenched state of a given merger tree need not be invariant across \textsc{Galacticus} parameter choices. We therefore treat quenched fractions and variability in the quenched state as data properties in their own right, and we will return to them explicitly in Section~\ref{sec:results}.

\paragraph*{Floor defined gas metal mass label:}\ 
A similar feature appears in gas metal mass, where a non-negligible fraction of systems lie exactly at the imposed floor value in log space. We therefore define a corresponding binary label that marks systems at the gas metal mass floor. Together, the quenched label and the floor state label provide compact summaries of the low-\sSFR{} and floor dominated regimes in the target distributions.

\subsection{Sample Construction and Data Split}\label{sec:data_split}
Given the resolution limit of the simulation \citep{ishiyama2021uchuu, poole2017convergence}, we adopt another cut on the merger trees of the dark matter halos. In particular, we build the models on trees whose $z=0$ root halos satisfy
\[
\log_{10}\!\left(M_{\rm halo}/(h^{-1} M_\odot)\right) > 10.
\]
After this cut, 201,986 merger trees remain in the sample. For the main experiments, we reserve 20,000 trees for testing and use eight disjoint chunks of 20,000 trees each for training. {Note that these merger trees cover a wide range of halo masses; however, because the sample size is very small, they may not provide an accurate representation of the halo mass function. }

The full data set contains 7,800 \textsc{Galacticus} catalogs sampled in the 17-dimensional parameter space \footnote{The original mock design consists of 10,000 parameter points generated with a Latin-hypercube sampling scheme. The present analysis uses a working subset of 7,800 catalogs, chosen so that preprocessing, training, and evaluation could be carried out consistently across all experiments reported here.}
. Of these, 600 are reserved for testing, and the remaining 7,200 form the training pool. The training catalogs are partitioned into six SAM groups, each corresponding to 1,080 training catalogs after reserving 120 catalogs for validation. 

{{The six SAM-group partitions and eight training-tree chunks define a $6\times8$ grid of 48 training realizations, each pairing one SAM group with one tree chunk and yielding an independently trained model. In the following, $\mathrm{SXX}$ denotes a SAM-group identifier and $\mathrm{TXX}$ a tree-chunk identifier, where XX is the index. The six training SAM groups are labeled $\mathrm{S01}$--$\mathrm{S06}$, while $\mathrm{S00}$ is reserved for the common test set of 600 held-out SAM catalogs. Similarly, the eight training chunks are labeled $\mathrm{T01}$--$\mathrm{T08}$, while $\mathrm{T00}$ denotes the common held-out test chunk of 20,000 merger trees. We denote a GNN model trained on a particular SAM-group--tree-chunk pair by $\mathrm{GNN}_{\mathrm{SXX/TXX}}$; for example, $\mathrm{GNN}_{\mathrm{S04/T04}}$ is the single-model realization trained on SAM group S04 and tree chunk T04. Unless otherwise stated, the single-model results in Sections~\ref{sec:results_stellar_mass} and \ref{sec:results_other_properties} use $\mathrm{GNN}_{\mathrm{S04/T04}}$ as a fixed reference realization. This fixed choice provides a consistent single-model reference, but it is not assumed to represent typical performance across the 48 realizations. We also construct a 48-model ensemble by averaging the predictions of these same models before computing the evaluation metrics. Appendix~\ref{app:realization_variation} presents the variation in single-model performance across the 48 realizations.}}

Within this {halo-mass selected} population, the tree split is held fixed across all model variants in the main text. The resulting evaluation, therefore, tests generalization along two axes simultaneously: to merger trees in a held-out tree split and to \textsc{Galacticus} catalogs in the 600-catalog test set, both excluded from training.

\section{Method}\label{sec:methodology}
We seek a model that maps halo merger trees and a SAM parameter vector to galaxy properties at the nine redshifts listed in Section~\ref{sec:targets}. A neural network is a function with a large number of free parameters (weights) that are adjusted during training to minimize the difference between the model's predictions and the target values. Here we use a graph neural network, a variant designed for data that are naturally represented as a network of connected nodes. In our case, halos are linked by merger and host--satellite relations. The key idea is that a halo's galaxy properties depend not only on its own mass and concentration, but also on its assembly history and environment. The network is conditional in the sense that it takes both the merger tree graph and the SAM parameter vector as input, so the same halo can yield different galaxy predictions under different baryonic prescriptions.

\subsection{Graph Neural Network Backbone}\label{sec:gnn}
We represent the halo population associated with a set of merger trees as a graph \(\mathcal{G}=(\mathcal{V},\mathcal{E})\). The graph has three components:
\begin{itemize}
    \item \textbf{Nodes.} Each node \(x\in\mathcal{V}\) denotes a halo and is assigned the feature vector \(\mathbf{x}_{\rm halo}\) defined in Equation (\ref{eq:halo_features}), which contains several fundamental halo properties together with time information.

    \item \textbf{Edges.} The edge set contains both progenitor--descendant links and host--satellite links. The former connect halos across simulation outputs along the merger history, while the latter connect subhalos to their hosts within the same output. In the present implementation, these edge types are used to define the graph connectivity along which information is propagated, but no additional edge features are assigned to encode properties of these relations.
    \item \textbf{Graph-level features.} Each \textsc{Galacticus} catalog is associated with a SAM parameter vector \(\mathbf{u}\), which supplies the global conditioning information for the model. Concretely, \(\mathbf{u}\) specifies the baryonic parameter setting of the catalog and is used to condition the predictions for all trees drawn from that catalog.
\end{itemize}
This construction allows the model to propagate information along the assembly history of a halo and between host halos and their satellites at fixed simulation output.

In practice, the original UchuuMicro merger trees contain 50 snapshots, but we do not use all of them when constructing the graph. To reduce the data volume while retaining the main temporal structure of the trees, we subsample the merger history to 15 predefined snapshot times between \(z=0\) and \(z=5.15\). These 15 snapshots include the nine redshifts at which galaxy targets are supervised, together with additional intermediate snapshots that provide a denser tracing of the merger history. Along each merger tree, we follow the main progenitor branch and connect the selected halos with progenitor--descendant edges between successive traced times; host--satellite edges are then added within each snapshot. The resulting graph is therefore temporally denser than the target supervision itself, while remaining substantially smaller than the full 50-snapshot tree representation. We do not train directly on the full graph; instead, as described below, each optimization step uses an induced subgraph built from the merger trees associated with the target galaxies in the batch.

\begin{figure*}
  \centering
  \includegraphics[width=0.96\textwidth]{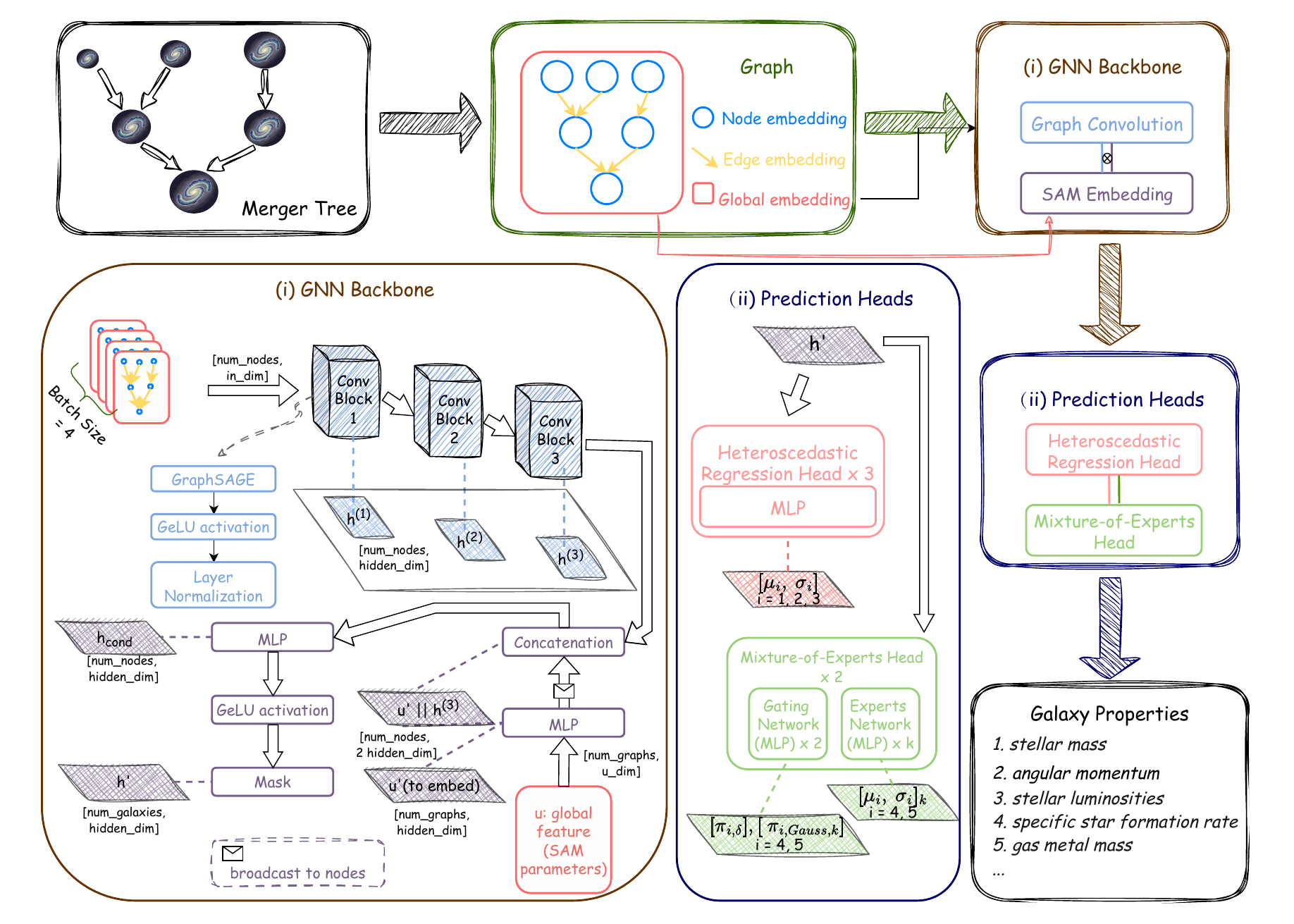}
  \caption{Schematic overview of the model pipeline. Halo merger trees are encoded by a shared GNN backbone, conditioned on the SAM parameter vector for the catalog, and evaluated with separate prediction heads for each redshift. In the current architecture, \(\Mstar\), \(\Lz\) and \(\Jang\) are modeled with heteroscedastic regression heads, while \sSFR\ and \(\Mzgas\) are modeled with mixture-of-experts heads.}
  \label{fig:pipeline}
\end{figure*}

Figure~\ref{fig:pipeline} summarizes the overall architecture. A shared GNN backbone, i.e., the common graph encoder used for all targets and redshifts, first maps the input graph into {latent halo representations} \footnote{{The graph encoder refers to the shared part of the network that transforms the input halo graph into latent representations, i.e., internal feature vectors learned for each halo that summarize information useful for predicting galaxy properties. These internal vectors are in the hidden space, which corresponds to the output $h^{(i)}$ of blue blocks in the GNN Backbone part of Figure~\ref{fig:pipeline}.}}. These latent representations are then passed through conditioning modules, which incorporate the SAM parameter vector so that the same merger tree structure can yield different predictions under different SAM parameter settings. The conditioned representations are finally fed into separate prediction heads \footnote{The head refers to the final layer of a shared backbone, designed to produce the desired output for a specific task.
} to output the final predictive quantities, or the parameters of the predictive distribution, for each target at a given redshift.

Let \(\mathbf{h}_x^{(\ell)}\) denote the latent representation of halo \(x\) at layer \(\ell\). Following the GraphSAGE formalism \citep{hamilton2017inductive}, a widely used GNN scheme in which each node updates its representation by combining its own features with aggregated information from neighboring nodes, the neighborhood information is first aggregated as
\begin{equation}
\mathbf{m}_{x}^{(\ell)}=
\operatorname{AGG}^{(\ell)}
\left(
\left\{
\mathbf{h}_{n}^{(\ell)}:n\in\mathcal{N}(x)
\right\}
\right),
\end{equation}
where \(\mathcal{N}(x)\) is the neighborhood of halo \(x\). and in our implementation 
 $\operatorname{AGG}$ denotes mean aggregation over neighboring halos. The node state is then updated through
\begin{equation}
\mathbf{h}_{x}^{(\ell+1)}=
\sigma\!\left(
\mathbf{W}^{(\ell)}
\left[
\mathbf{h}_{x}^{(\ell)} \,\Vert\, \mathbf{m}_{x}^{(\ell)}
\right]
\right),
\end{equation}
where \(\mathbf{W}^{(\ell)}\) is a learnable weight matrix, \(\Vert\) denotes concatenation, and \(\sigma\) {is a nonlinear activation function, taken here to be Gaussian Error Linear Units (GELUs; \citealt{hendrycks2016gelu})}. In other words, each layer updates the representation of a halo by combining its current state with summary information from the halos connected to it in the graph. In practice, we use a GraphSAGE backbone because it is inductive (i.e., it can generalize to completely new, unseen merger trees without retraining), stable for large graphs, and well-suited to the irregular structure of merger trees. The same GELU activation is used throughout the multilayer perceptron (MLP) blocks in the GNN backbone, conditioning modules, and output heads.

\subsection{SAM Conditioning}
The full halo graph described in Section~\ref{sec:data} is too large to process in a single optimization step, so the training is carried out on induced subgraphs associated with selected merger trees. For a given batch, we first identify the trees contributing the target galaxies and then extract the corresponding halo nodes and edges. Each subgraph, therefore, retains the relevant merger history and host-satellite structure while remaining computationally manageable.

Each halo node is assigned the input vector \(\mathbf{x}_{\rm halo}\) defined in Equation (\ref{eq:halo_features}). Message passing (the iterative process by which each halo node aggregates information from its graph neighbors, as described in Section~\ref{sec:gnn}) is performed on a single adjacency that contains both relation types, i.e., the spatial relations of halos and connections between progenitor and descendants; we do not introduce separate specific convolution weights for progenitor--descendant and host--satellite edges. After the shared GNN backbone, each halo is represented by a latent representation \(\tilde{\mathbf{h}}_x \in \mathbb{R}^{d}\), where \(d\) is the hidden dimension, i.e., the size of the internal feature vector used by the network.

The same merger tree can produce different galaxy populations when evolved with different SAM parameter choices. To capture this dependence, we condition the latent halo representation produced by the graph encoder on the SAM parameter vector \(\mathbf{u}\) of the corresponding catalog. {We first map the SAM parameter vector \(\mathbf{u}\in\mathbb{R}^{17}\) into the same hidden dimension and then combine it with the halo representation through
\begin{equation}
\tilde{\mathbf{h}}_x^{\,\mathrm{cond}}
=
\sigma\!\left(
\mathbf{W}_{\mathrm{cond}}
\left[
\tilde{\mathbf{h}}_x \,\Vert\, (\mathbf{W}_{u}\mathbf{u} + \mathbf{b}_{u})
\right]
\right),
\end{equation}
where \(\mathbf{W}_{u}\mathbf{u} + \mathbf{b}_{u}\in\mathbb{R}^{d}\) is a learned linear projection of the SAM parameter vector into the hidden dimension, \(\mathbf{W}_{\mathrm{cond}}\) is a learned fusion matrix, and \(\sigma\) denotes the same nonlinear activation used elsewhere in the network.} This provides a simple and stable way to let the same graph encoder respond differently to different semi-analytic prescriptions without changing the graph itself.

The galaxy targets are sparse on this graph: not every halo node hosts a target galaxy at every redshift. We therefore supervise the model only on the subset of halo nodes associated with \textsc{Galacticus} galaxies, together with a redshift index \(t\) indicating which snapshot each target belongs to. If a target galaxy at redshift \(z_t\) is associated with halo \(x\), we denote its conditional latent representation by \(\tilde{\mathbf{h}}_{x,t}^{\,\mathrm{cond}}\). This allows one merger tree subgraph to contribute supervision at multiple snapshots without requiring dense labels on all halos.

\subsection{Prediction Heads}
Predictions are made only at the halo nodes associated with target galaxies. For each redshift \(z_t\), the corresponding conditional latent representations are passed to a head for that redshift, yielding one set of predictions for each of the five target properties at that snapshot. This gives the model a shared graph encoder, but allows the final mapping from latent halo state to galaxy properties to vary with redshift. The prediction heads are therefore designed to produce not only point estimates, but also distributional outputs, represented either by a mean prediction together with its expected scatter or by a full conditional distribution, depending on the target.

For \(\Mstar\), \(\Lz\) and \(\Jang\), we use heteroscedastic regression heads that predict both a conditional mean \(\mu\) and a conditional scatter \(\sigma\), allowing the expected spread around the mean prediction to vary with the input halo and SAM  \citep{nix1994estimating, jespersen2022mangrove}. These heads are used for the smooth targets, for which a single conditional distribution remains a reasonable approximation, while still allowing the predictive uncertainty to vary across the input space. For \sSFR\ and \(\Mzgas\), in order to represent the pronounced low-\sSFR\ branch and floor-dominated structure present in these targets, we use mixture-of-experts (MoE) heads \citep{jacobs1991adaptive} which represent the target distribution as a weighted combination of several simple components rather than a single Gaussian. The detailed training objectives for these heads are described in Section~\ref{sec:training}.

When a target is modeled with a mixture head, evaluation requires a single point estimate derived from the predicted distribution. Unless otherwise stated, the reported tables and figures use the maximum-a-posteriori (MAP) component as that point summary.

\section{Training}\label{sec:training}
We train the model with a joint objective over the nine redshift outputs. Throughout training, halo node features, SAM parameters, and galaxy targets are standardized using the mean and variance computed from the training data only. This places the five targets on comparable numerical scales while preserving the sparse supervision at each redshift described in Section \ref{sec:methodology}.

\subsection{Multi-task Loss}
For the first three targets, \(\Mstar\), \(\Lz\), and \(\Jang\), the network predicts both a mean and a logarithmic scatter term, and we optimize a heteroscedastic Gaussian negative log-likelihood in normalized target space,
\begin{equation}
\mathcal{L}_{\mathrm{het}}
= \frac{1}{N}\sum_{i=1}^{N}
\left[
-\log \mathcal{N}(y_i \mid \mu_i, \sigma_i^2)
\right].
\end{equation}
{{Here \(i\) indexes the galaxies with a valid value of the target at the
redshift under consideration, and \(N\) is the number of such galaxies in the
current training batch.}} The quantity
\(\mathcal{N}(y_i\mid\mu_i,\sigma_i^2)\) denotes the Gaussian likelihood with
predicted mean \(\mu_i\) and variance \(\sigma_i^2\).
This allows the model to adapt the effective regression weight to the local predictive uncertainty while retaining a single shared backbone.

For \(\sSFR\) and \(\Mzgas\), the output distributions contain a substantial accumulation at the quenched or floor-defined branch. We therefore use mixture heads with one narrow floor component and several continuous Gaussian components. For a normalized target \(y\), the corresponding negative log-likelihood is
\begin{equation}
\mathcal{L}_{\mathrm{mix}}
= -\log \left[
\pi_{0}\,\mathcal{N}(y\,|\,y_{\mathrm{floor}},\sigma_{0}^{2})
\;+\;
\sum_{k=1}^{K}\pi_{k}\,\mathcal{N}(y\,|\,\mu_{k},\sigma_{k}^{2})
\right].
\end{equation}
{{Here \(K\) is the number of continuous Gaussian components, \(\pi_0\) is the
predicted mixture weight of the quenched/floor-state component, and \(\pi_k\) is the predicted
weight of the \(k\)th continuous component. The weights are non-negative and
satisfy
\(\pi_0+\sum_{k=1}^{K}\pi_k=1\), while \(\mu_k\) and \(\sigma_k\) are the
predicted mean and standard deviation of the \(k\)th continuous component.}}
The value \(y_{\mathrm{floor}}\) is fixed to the quenched threshold for
\(\sSFR\) or the adopted floor value for \(\Mzgas\). In practice, values at or below these thresholds are snapped to the corresponding floor branch during mixture training so that the point mass component learns the discrete accumulation explicitly.

The total objective is the sum of the losses at each redshift over all available labels in the batch,
\begin{equation}
\mathcal{L}_{\mathrm{tot}}
=
\sum_{t=1}^{9}
\left(
\mathcal{L}_{\mathrm{het},t}
\;+\;
\mathcal{L}_{\mathrm{mix},t}
\right),
\end{equation}
averaged over the number of supervised targets contributing at each redshift.

\subsection{Optimization}
The main single model is trained with the AdamW optimizer \citep{loshchilov2019decoupled} using a learning rate of \(5\times10^{-4}\) and weight decay \(10^{-4}\), chosen on the basis of preliminary validation experiments. Because each training example is a large merged subgraph, we use a batch size of one catalog and accumulate gradients over 32 updates before each optimizer step. We train for at most 350 epochs and monitor the validation loss with a learning rate schedule based on validation plateaus (reduction factor 0.2, scheduler patience 8) together with early stopping after 10 non-improving epochs. The experiments reported here were run on a single NVIDIA GeForce RTX 4090 GPU with 24\,GB of memory.

\subsection{Evaluation}
For \(\sSFR\) and \(\Mzgas\), the target scaling is determined from the continuous branch so that the discrete floor population does not dominate the mean and variance. The checkpoint with the lowest overall validation loss is retained and then evaluated on the fixed test catalogs and fixed test trees defined in Section~\ref{sec:data_split}.

For each target property and redshift snapshot, we report the bias,
\begin{equation}
\mathrm{bias} = \mathbb{E}[\hat{y}-y],
\end{equation}
the scatter,
\begin{equation}
\sigma = \mathrm{Std}(\hat{y}-y),
\end{equation}
the Pearson correlation coefficient,
\begin{equation}
\rho = \frac{\mathrm{Cov}(y,\hat{y})}{\sigma_y \sigma_{\hat{y}}},
\end{equation}
and the coefficient of determination,
\begin{equation}
R^2 = 1 - \frac{\sum_i (y_i-\hat{y}_i)^2}{\sum_i (y_i-\bar{y})^2},
\end{equation}
where \(y\) and \(\hat{y}\) denote the true and predicted values in \(\log_{10}\) units, and \(\bar{y}\) is the sample mean of the true values.

For \(\sSFR\), we also evaluate how well the model separates quenched and star forming galaxies using the binary definition introduced in Section~\ref{sec：classification_label}. For \(\Mzgas\), we evaluate the corresponding classification problem for galaxies that lie at the adopted floor value, \(\log(M_{Z,\mathrm{gas}})=-5\). In both cases, the classification results reported below use precision and recall,
\begin{equation}
\mathrm{precision} = \frac{\mathrm{TP}}{\mathrm{TP}+\mathrm{FP}},
\qquad
\mathrm{recall} = \frac{\mathrm{TP}}{\mathrm{TP}+\mathrm{FN}},
\end{equation}
where \(\mathrm{TP}\), \(\mathrm{FP}\), and \(\mathrm{FN}\) denote the numbers of true positives, false positives, and false negatives, respectively,
and summarize them with the \(F_1\) score,
\begin{equation}
F_1 = \frac{2\,\mathrm{precision}\times\mathrm{recall}}{\mathrm{precision}+\mathrm{recall}},
\end{equation}
which summarizes the balance between purity and completeness for the positive class. Regression metrics for \(\sSFR\) are computed only on the continuous branch, \(\logsSFR > \ssfrquenchthr\), so that the quality of the continuous prediction is not conflated with the separate problem of identifying the quenched population. Unless stated otherwise, all tabulated metrics are pooled over galaxies across the full test set, while plotting panels use random subsamples for readability.

\section{Results}\label{sec:results}

In this section, we summarize the predictive performance of the surrogate model for the five target galaxy properties across the nine redshift outputs. Unless otherwise stated, {{the results in
Sections~\ref{sec:results_stellar_mass} and \ref{sec:results_other_properties} refer to the fixed reference realization $\mathrm{GNN}_{\mathrm{S04/T04}}$ defined in Section~\ref{sec:data_split}, evaluated on the common $\mathrm{S00/T00}$ test set.}} We first present {{this reference model}}, beginning with stellar mass as the cleanest reference case, then extend the comparison to the remaining targets, and finally summarize the effect of ensemble averaging.

\subsection{Stellar Mass Results}\label{sec:results_stellar_mass}
\begin{figure}
  \centering
  \includegraphics[width=\columnwidth]{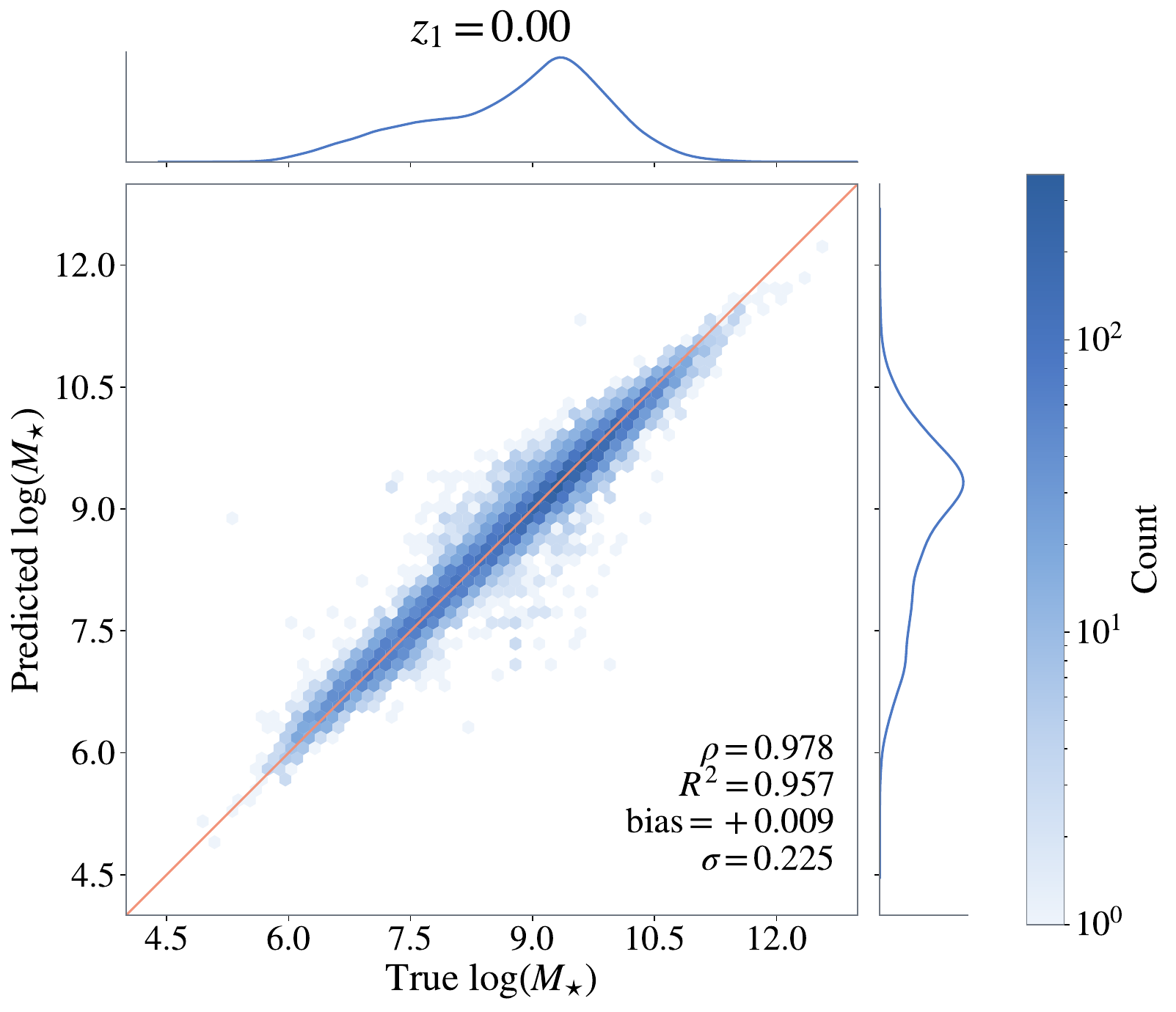}
  \caption{Predicted versus true stellar masses at $z=0$. The one-to-one line is shown in orange, while the upper and right panels show the corresponding true and predicted marginal distributions. The inset reports the metrics for the main panel.}
  \label{fig:fig01}
\end{figure}
We begin with stellar mass, which provides a benchmark for whether the model captures the mapping from halo assembly history and SAM parameters to galaxy properties. At $z=0$, our GNN model yields a scatter of 0.225, a bias of $0.009$, a correlation coefficient of 0.978, and $R^2=0.957$ for \(\log(M_\star)\) as shown in Figure~\ref{fig:fig01}. The predicted and true stellar masses follow a tight one-to-one relation over most of the dynamic range, indicating that stellar mass is one of the most accurately recovered targets in the current setup.
\begin{figure}
  \centering\includegraphics[width=\columnwidth]{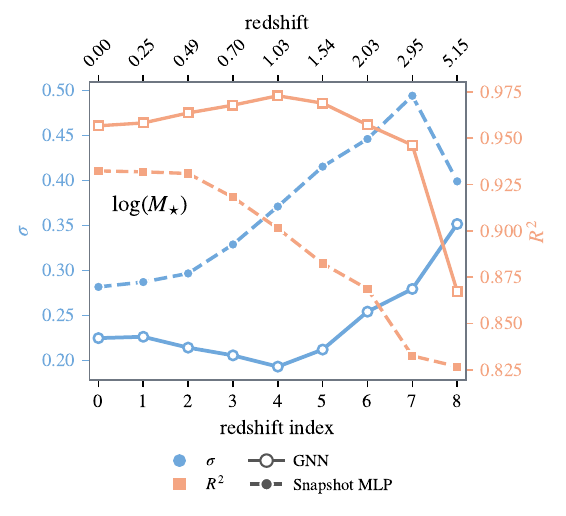}
  \caption{{{Stellar mass performance as a function of redshift. Solid curves with open markers show the reference GNN, while dashed curves with filled markers show independently trained single snapshot MLPs at the same redshift. Blue circles denote the scatter, and orange squares denote $R^2$.}}}
  \label{fig:fig03}
\end{figure}
\paragraph*{$z=0$ baseline.}\ 
{{As a baseline model, we train an MLP that uses only the halo properties
measured at the target snapshot together with the SAM parameter vector. It
does not receive progenitor and descendant connections or halo states from
other snapshots. At $z=0$, the MLP uses the S04/T04 training objects and is
evaluated on S00/T00, matching the catalog and object selections used for the
reference GNN. Here T04 and T00 identify the merger tree chunks from which the
individual objects are selected; the MLP itself receives no merger tree
structure. Its $z=0$ stellar mass metrics, reported in
Table~\ref{tab:z0_regression_summary}, are weaker than those of the reference
GNN: the MLP gives a scatter of $0.279$, a bias of $0.008$,
$\rho=0.967$, and $R^2=0.933$, compared with a scatter of $0.225$ and
$R^2=0.957$ for the GNN.}} This comparison indicates that explicit assembly history information improves stellar mass prediction beyond what can be inferred from single snapshot halo properties alone.

\paragraph*{Results across redshift.}\ The stellar mass prediction remains stable beyond the first snapshot. Figure~\ref{fig:fig03} summarizes the evolution of the stellar mass scatter and $R^2$ over the nine redshift outputs. Across the full redshift range, the stellar mass bias stays small, from $0.009$ at low redshift to $-0.030$ at the highest redshift output. The scatter remains near 0.19--0.28 through the intermediate snapshots before increasing to 0.352 at the highest redshift, while $R^2$ stays between 0.946 and 0.973 through the first eight outputs and decreases to 0.867 only at the final snapshot. 

{{To extend the baseline comparison beyond $z=0$, we independently train the
same MLP architecture at each of the other eight snapshot outputs, retaining
the S04/T04 training and S00/T00 evaluation selections. The GNN has lower scatter and higher
$R^2$ than the corresponding MLP at every output, as expected. The reduction in scatter is
about $20$--$21\%$ at $z=0$--$0.25$, increases to $48$--$49\%$ at
$z=1.03$--$1.54$, remains approximately $43\%$ at $z=2.03$--$2.95$, and is
$12\%$ at $z=5.15$. Across the nine outputs, the corresponding increase in
$R^2$ ranges from $0.024$ to $0.114$. The consistent advantage of the GNN therefore supports
the conclusion that halo assembly information provides predictive value
beyond the instantaneous halo state, with the difference becoming
particularly clear over $z\simeq1$--$3$.
}}

Stellar mass remains one of the most stable targets in our set, consistent with its character as an integrated quantity that depends on the cumulative growth history of the system. The degradation toward the highest redshift output is also consistent with the increasingly sparse galaxy population and the reduced amount of accumulated assembly information available in the trees. Even so, $\Mstar$ remains accurately predicted over most of the sampled redshift range and therefore provides a useful reference point for interpreting the more difficult targets below. 
\begin{figure*}[t]
  \centering
  \includegraphics[width=0.98\textwidth,keepaspectratio]{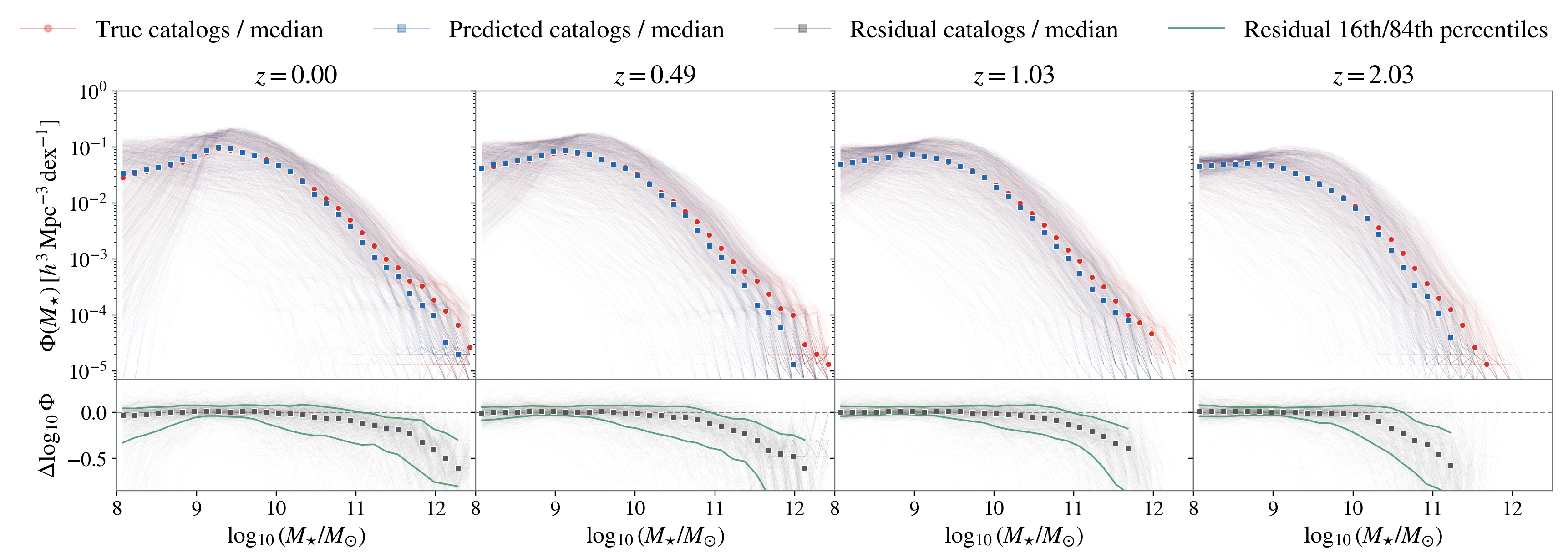}
  \caption{{Stellar mass functions of the 600 test SAM catalogs at four redshift outputs. Faint red and blue curves show the true and predicted functions of individual catalogs, and the corresponding markers show their medians. The flattening and decline at the lower mass end is simply caused by the resolution and incompleteness of the simulation. The lower panels show $\Delta\log_{10}\Phi$ for individual catalogs, with gray squares and green curves denoting the median and the 16th and 84th percentiles, respectively.}}
  \label{fig:smf_single_sam}
\end{figure*}
\paragraph*{Population-level diagnostic.}\ 
{{Figures~\ref{fig:smf_single_sam} and \ref{fig:2pcf_single_sam} provides  population-level diagnostics through the stellar mass function and two-point correlation function. For these diagnostics, each of the 600 held-out SAM catalogs is evaluated over
the full mass-selected set of 201,986 merger trees described in
Section~\ref{sec:data_split}. Figure~\ref{fig:smf_single_sam} shows that the
median predicted stellar mass function follows the median true abundance over
the low- and intermediate-mass ranges.}} The lower row summarizes the residuals across the 600 test SAM catalogs. This makes clear that the agreement is not uniform: the largest discrepancies appear toward the high mass end, where the model tends to underpredict the abundance of the rarest systems, and this tendency becomes more pronounced at higher redshift. 

This behaviour is broadly consistent with the pointwise results above and the small simulation volume we use, which suggests there are not enough massive halos to yield a stable estimate insensitive to sample variance. Most galaxies still lie close to the one-to-one relation in Figure~\ref{fig:fig01}, so the global scatter and $R^2$ remain strong, but the stellar mass function can be affected specifically by the small number of objects in the high mass tail. A modest negative bias for those rare systems, therefore, can produce a much more visible deficit in the abundance than in the global regression metrics alone.
\begin{figure*}
  \centering
  \includegraphics[width=0.98\textwidth,keepaspectratio]{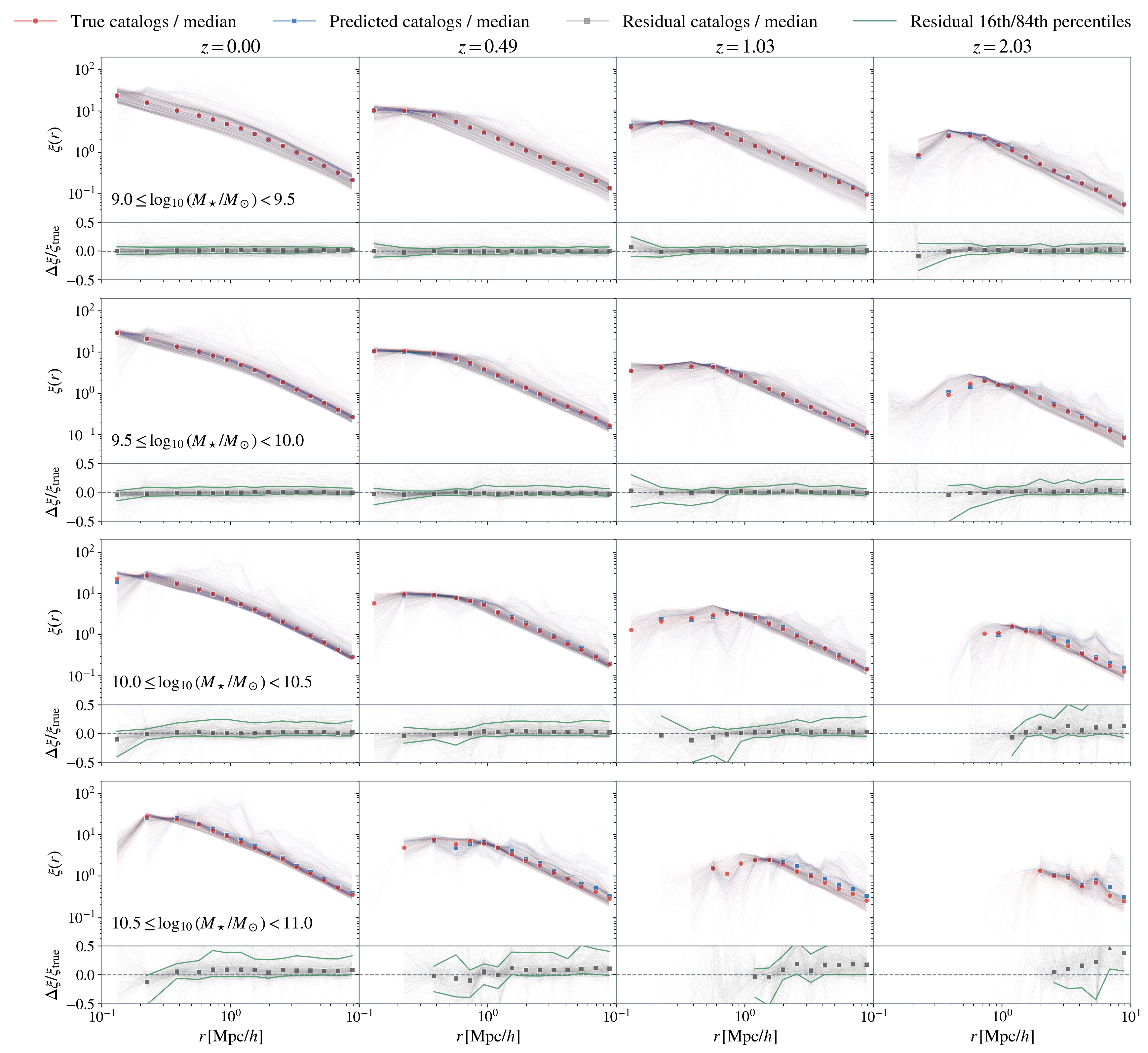}
  \caption{{{Two-point correlation functions across four redshift outputs. Each pair of rows corresponds to one stellar mass bin: the upper panels compare the predicted and true functions for the 600 test SAM catalogs, while the lower panels show the fractional residuals and their median and 16th--84th percentile range. The agreement is generally good over the intermediate scales and mass ranges. The largest deviations occur in the highest-mass and highest-redshift bins, where the galaxy sample is sparse. In addition, the limited simulation volume ($100\,\mathrm{Mpc}/h$ on a side) reduces the number of close galaxy pairs on small scales and the number of independent modes on large scales, making the clustering estimates noisier at both ends of the scale range. Nevertheless, the overall consistency between the SAM catalog and the GNN prediction demonstrates that the surrogate captures the main clustering behavior of the underlying model.}}}
  \label{fig:2pcf_single_sam}
\end{figure*}
Figure~\ref{fig:2pcf_single_sam} extends the comparison from abundance statistics to spatial clustering. {{The median predicted and true correlation functions agree
closely over intermediate scales in the two lower stellar mass bins.}} The discrepancies become more visible in the highest mass bins and at higher redshift, where the galaxy sample is sparse, and the clustering estimate is correspondingly noisier. 

\begin{figure*}
  \centering
  \includegraphics[width=0.98\textwidth,keepaspectratio]{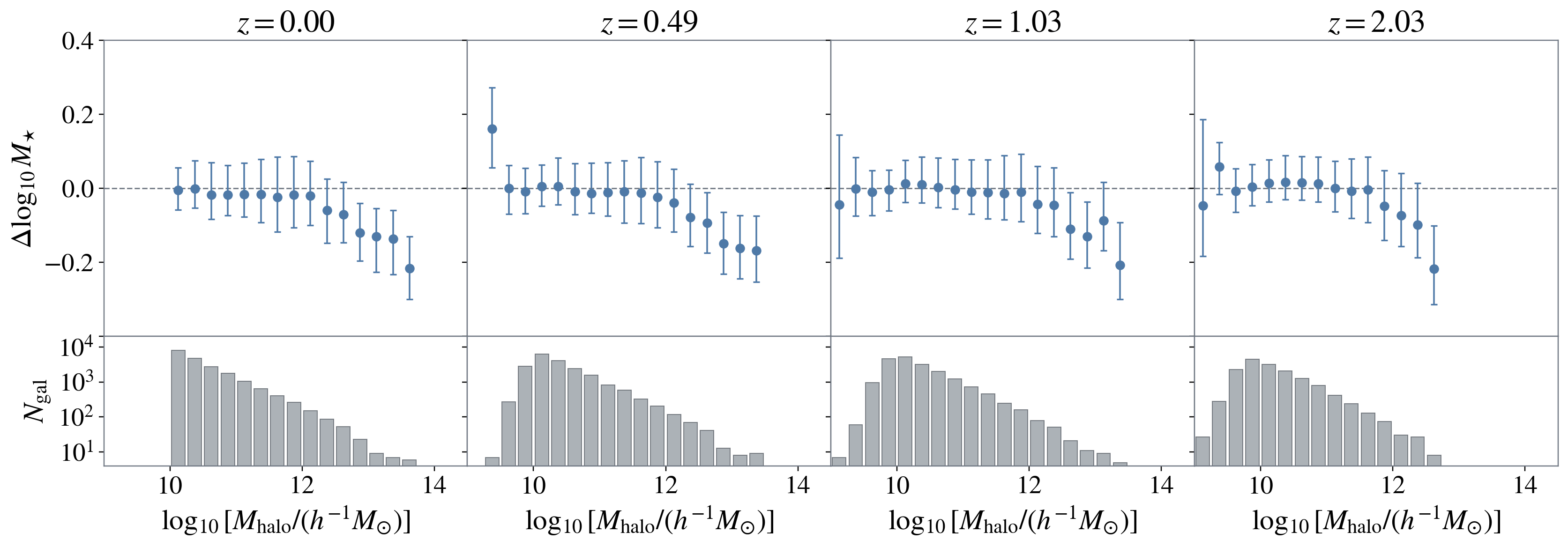}
  \caption{{{Residuals of the stellar mass prediction as a function of halo mass for $\mathrm{GNN}_{\mathrm{S04/T04}}\!\rightarrow\!\mathrm{S00/T00}$ at four redshifts. Points and error bars show the median and 16th--84th percentile range of the catalog-level median residuals in each halo mass bin. The lower panels show the median number of galaxies per catalog in the same bins.}}}
  \label{fig:mstar_residual_halo_mass}
\end{figure*}

{{Taken together, the two population diagnostics identify a common limitation in the rare, high-mass regime, which can become more pronounced toward higher redshift. To examine this discrepancy, we use the common T00 chunk of 20,000 merger trees, bin galaxies by their halo
mass at each redshift, and calculate the median stellar mass residual within
each test catalog. In Figure~\ref{fig:mstar_residual_halo_mass}, the median residual remains close to zero in bins containing many galaxies, but it becomes increasingly negative above approximately
$\log_{10}[M_{\rm halo}/(h^{-1}M_\odot)]\simeq12$ and reaches about
$-0.1$ to $-0.2$ dex in the highest bins. The lower panels show that the negative residuals occur where the number of galaxies declines by orders of magnitude. This implies that the finite simulation volume, and the resulting scarcity
of massive systems in the training set, can contribute to the failure mode. Appendix~\ref{app:halo_mass_catalogs} shows the corresponding 
residual distributions for five test catalogs.}}

\begin{figure*}
  \includegraphics[width=0.98\textwidth,keepaspectratio]{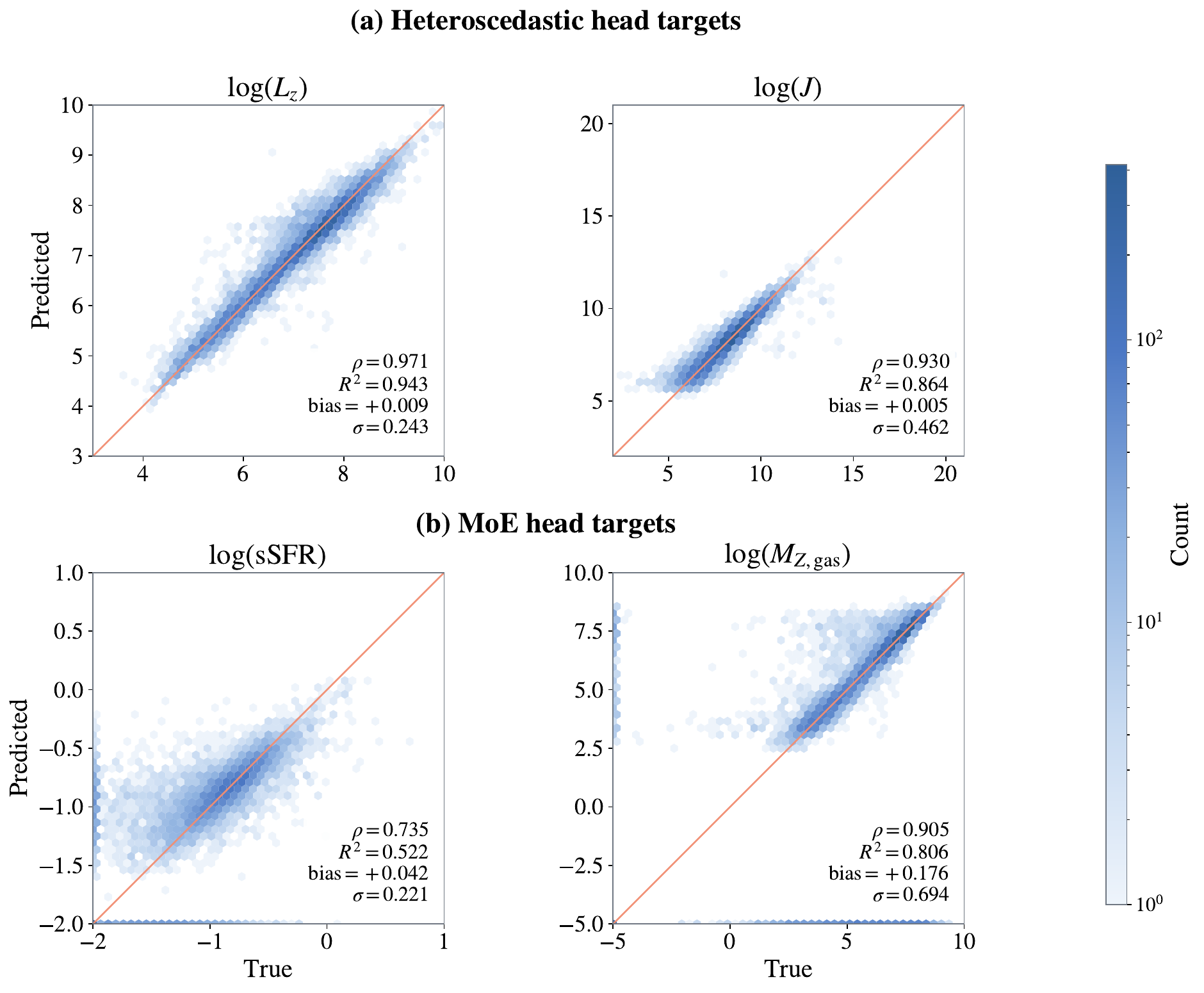}
\caption{Predicted versus true values at \(z_1=0\) for the four remaining targets. The upper row shows \(L_z\) and \(J\), and the lower row shows \(\mathrm{sSFR}\) and \(M_{Z,\mathrm{gas}}\). The one-to-one line is shown in orange in each panel.}
\label{fig:fig05}
\end{figure*}
\begin{figure*}
  \centering
  \includegraphics[width=0.98\textwidth]{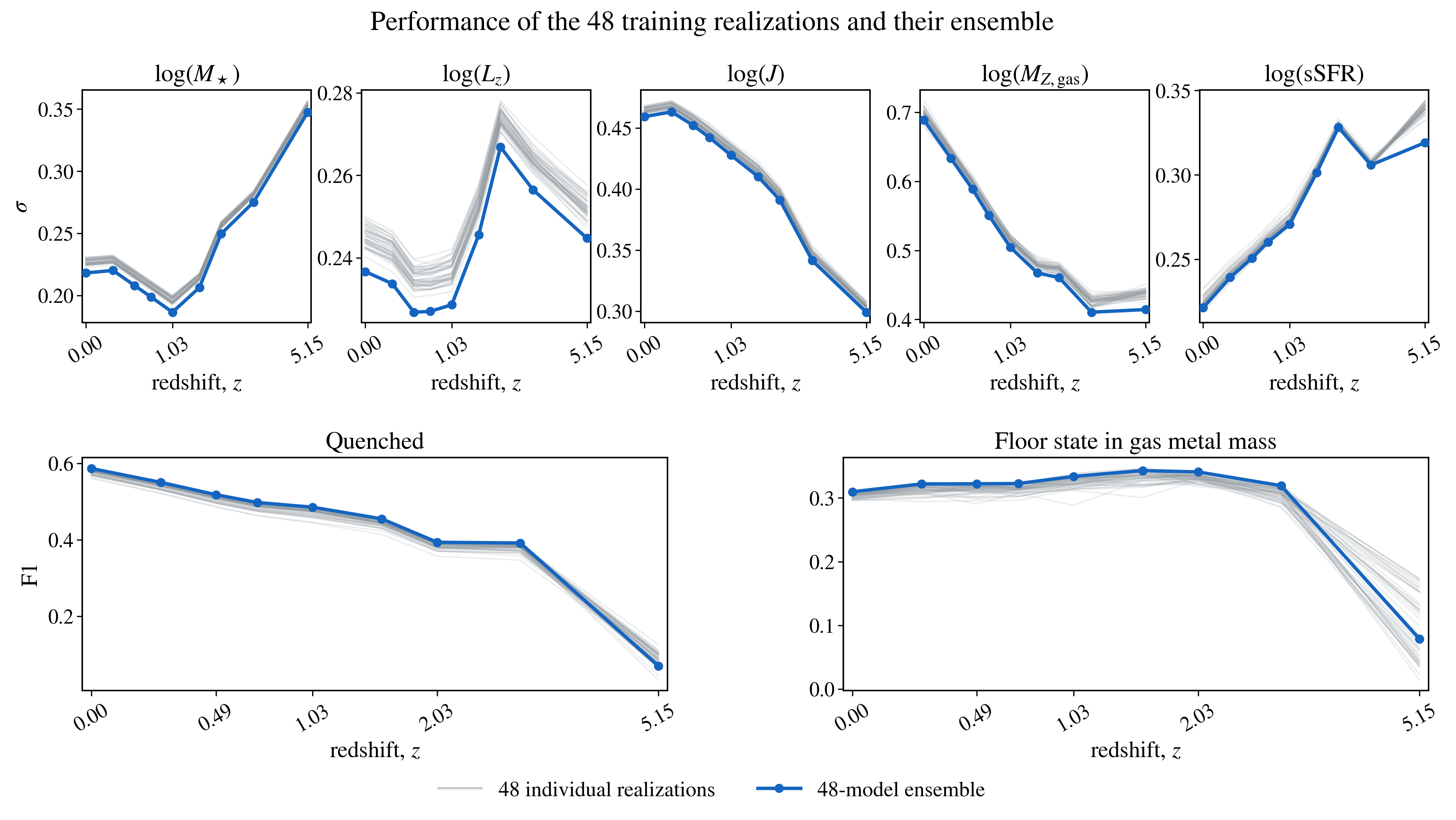}
  \caption{{{Performance of the models from the 48 training realizations and their ensemble across redshift. Each pale line shows one individual model, while the blue line shows the average of the 48-model ensemble. The upper panels report the continuous scatter $\sigma$; the lower panels report the $F_1$ score for the quenched state and the floor state in gas metal mass.}}}
  \label{fig:ensemble_performance}
\end{figure*}

\subsection{Other Galaxy Properties}\label{sec:results_other_properties}
As in the stellar mass case, the remaining targets are best interpreted relative to the same \(z=0\) MLP baseline summarized in Table~\ref{tab:z0_regression_summary}. These targets naturally separate into two groups according to their output behavior and head design. Luminosity and angular momentum are modeled with the same heteroscedastic regression head used for stellar mass, whereas \(\sSFR\) and \(\Mzgas\) use the MoE head because their conditional distributions contain distinct branches or floor populations. A broader visual summary of the predicted versus true relations across all five targets and several representative redshift outputs is provided in Appendix~\ref{Appendix:A} (Figure~\ref{fig:scatter_grid_overview}).

\paragraph*{Targets with heteroscedastic heads.}\  The luminosity and angular momentum results remain comparatively smooth across redshift, much like stellar mass. In the upper row of Figure~\ref{fig:fig05}, the predicted and true values for \(\Lz\) and \(\Jang\) stay close to the one-to-one relation at \(z=0\), and across the full redshift range their mean scatter values are 0.246 and 0.411, with corresponding mean \(R^2\) values of 0.940 and 0.884 respectively. Relative to the \(z=0\) MLP baseline in Table~\ref{tab:z0_regression_summary}, \(\Lz\) shows a modest but consistent improvement, with scatter decreasing from 0.269 to 0.243 and \(R^2\) increasing from 0.929 to 0.943. By contrast, \(\Jang\) is only marginally improved over the baseline at $z=0$. Luminosity remains a relatively smooth target for which information from the merger tree offers some additional leverage, whereas angular momentum appears harder to improve in this setting, presumably because its value is more sensitive to details of the dynamical history that are not fully captured by the present model, i.e., the current model only adopts the merger history of the dark matter halos, but the dynamics can also be affected by processes like smooth accretion which is not accounted for.

\paragraph*{Targets with MoE heads.}\ The more challenging behavior appears in \(\sSFR\) and \(\Mzgas\), which show the strongest non-Gaussian structure in the target set. The lower row of Figure~\ref{fig:fig05} makes this clear at \(z=0\): \(\sSFR\) contains a structured low \(\sSFR\) branch, while \(\Mzgas\) includes a prominent floor dominated population. For these two targets, the gain over the MLP baseline is substantially larger than for $\Lz$ or $\Jang$. For \(\Mzgas\), the mean scatter decreases from 1.233 to 0.536 and the mean \(R^2\) rises from 0.489 to 0.874; for \(\sSFR\), the mean scatter decreases from 0.335 to 0.280 and the mean \(R^2\) increases from 0.115 to 0.502. These targets, therefore, provide the clearest evidence that explicit tree information together with a MoE output head is most useful when the conditional distribution is not well described by a single smooth regression model.

\subsection{Ensemble Results}
In addition to {{the fixed reference realization $\mathrm{GNN}_{\mathrm{S04/T04}}$ reported above,}} we evaluate an ensemble model that averages predictions from models trained on different SAM groups and tree chunks. {{All models are evaluated on the common $\mathrm{S00/T00}$ test set, comprising the same 600 held-out SAM catalogs and the same unseen chunk of 20,000 merger trees. Figure~\ref{fig:ensemble_performance} shows the ensemble together with the individual models across redshift. Relative to $\mathrm{GNN}_{\mathrm{S04/T04}}$,}} the redshift-averaged regression metrics show a small but consistent improvement for most targets. For example, the mean stellar mass scatter decreases from 0.240 to 0.234, while the mean $R^2$ increases from 0.951 to 0.953.

The classification performance at $z=0$ remains broadly unchanged. For \(\sSFR\), $F_1$ changes from 0.588 to 0.586, wheras for \(\Mzgas\), $F_1$ changes from 0.303 to 0.310. This suggests that ensemble averaging mainly improves the stability of the continuous regression outputs, while it does not substantially change the discrete state identification performance in the current setup.

\section{Discussion}\label{sec:discussion}
\begin{figure*}
  \centering\includegraphics[width=0.96\textwidth]{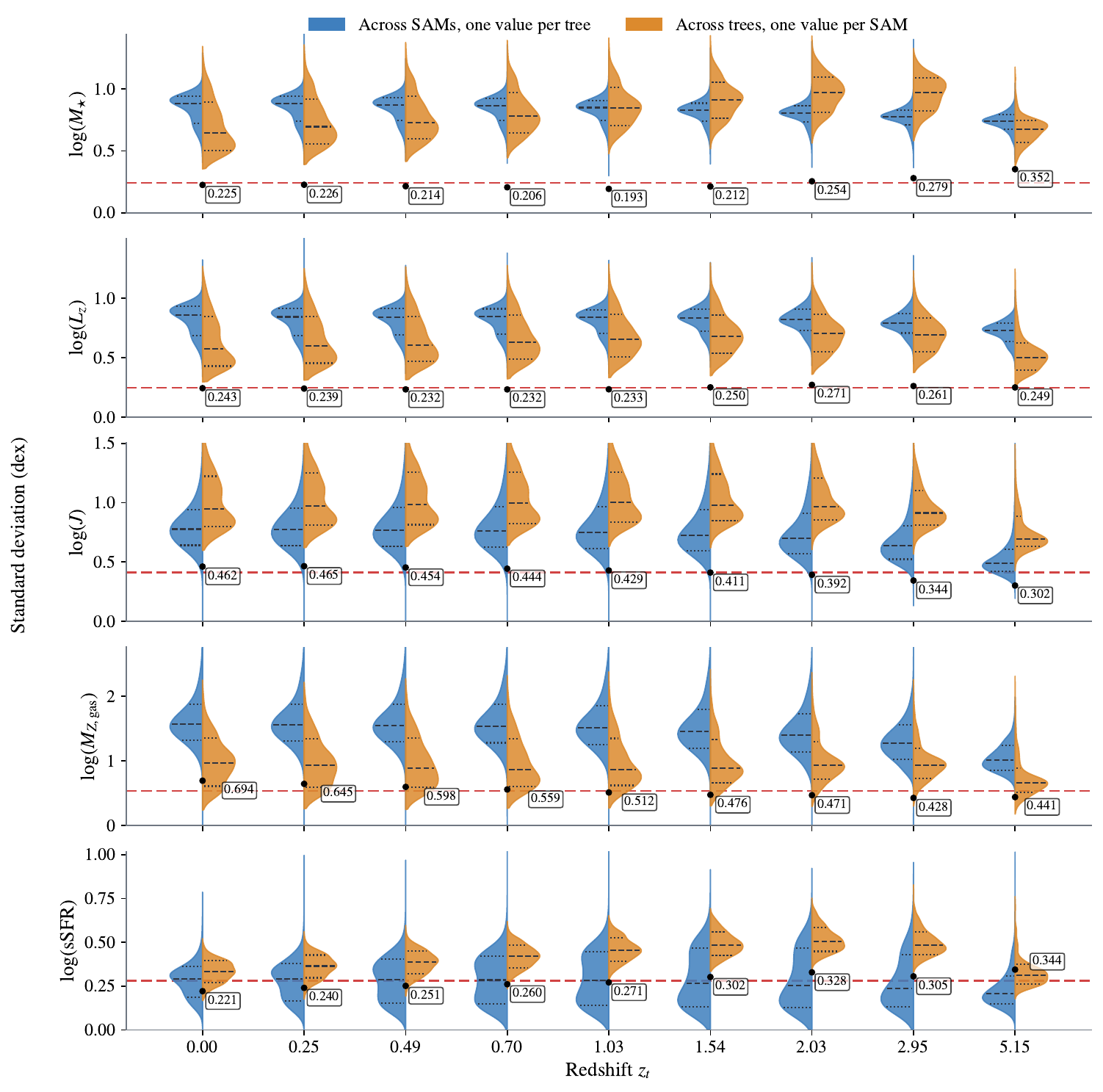}
  \caption{Continuous scatter distributions for the five target properties across the nine redshift outputs. In each row, the left half violin shows the distribution, over 20\,000 fixed test trees, of the standard deviation measured across 600 test SAM catalogs. The right half violin shows the distribution, over those 600 test SAM catalogs, of the standard deviation measured across the same fixed tree set. The horizontal reference line marks the prediction scatter for that property on the test set averaged over redshift, while points and numeric labels mark the corresponding prediction scatter at each redshift. For \(\sSFR\), quenched systems are excluded from the continuous summary; for \(M_{Z,\mathrm{gas}}\), systems at the floor branch are excluded.}
  \label{fig:sam_tree_variation_dumbbell}
\end{figure*}

We have presented a GNN-based model that can yield accurate and fast predictions of galaxy properties with the input of merger trees of dark matter halos and a SAM parameter set. We  note that the overall model performance can be affected by multiple factors, and the structure itself is worth further investigation. In this section, we present additional tests that can help us better understand the model.
\begin{figure}
  \centering
  \includegraphics[width=\columnwidth]{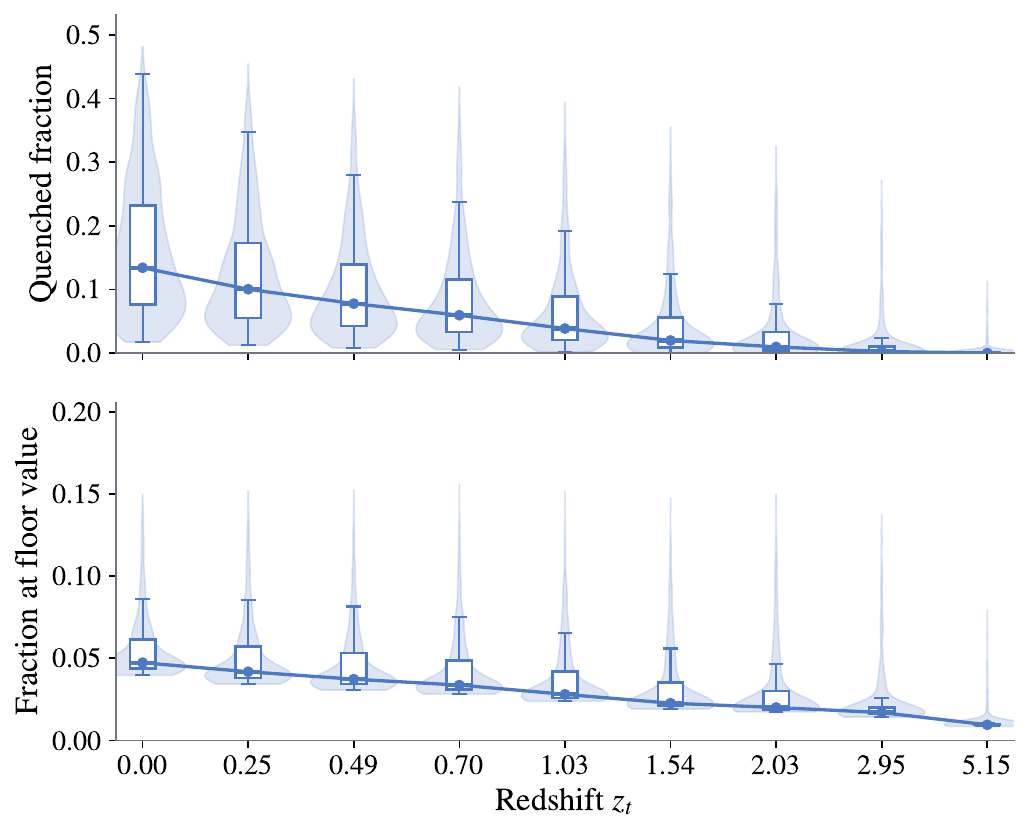}
  \caption{Catalog-level fractions for the mass-selected sample as a function of redshift. The upper panel shows the quenched fraction and the lower panel shows the fraction of systems at the floor value in gas metal mass. At each redshift, the violin indicates the full distribution across catalogs, the box marks the interquartile range, and the central line marks the median.}
  \label{fig:fig04}
\end{figure}

\begin{figure}
  \centering
  \includegraphics[width=\columnwidth]{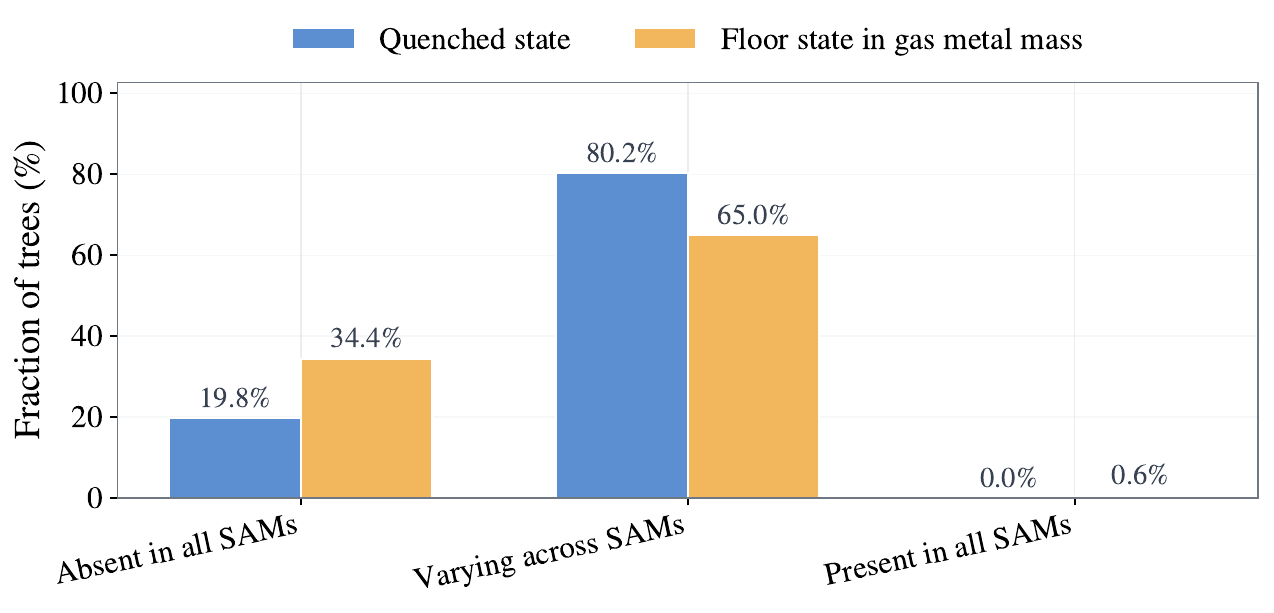}
  \caption{Fractions of test trees that are absent from a given state in all SAM catalogs, present in that state in all catalogs, or vary across catalogs at \(z=0\). Blue bars show the quenched state, and orange bars show the floor state in gas metal mass.}
  \label{fig:tree_state_variability}
\end{figure}

\begin{figure*}
  \centering
  \includegraphics[width=\textwidth]{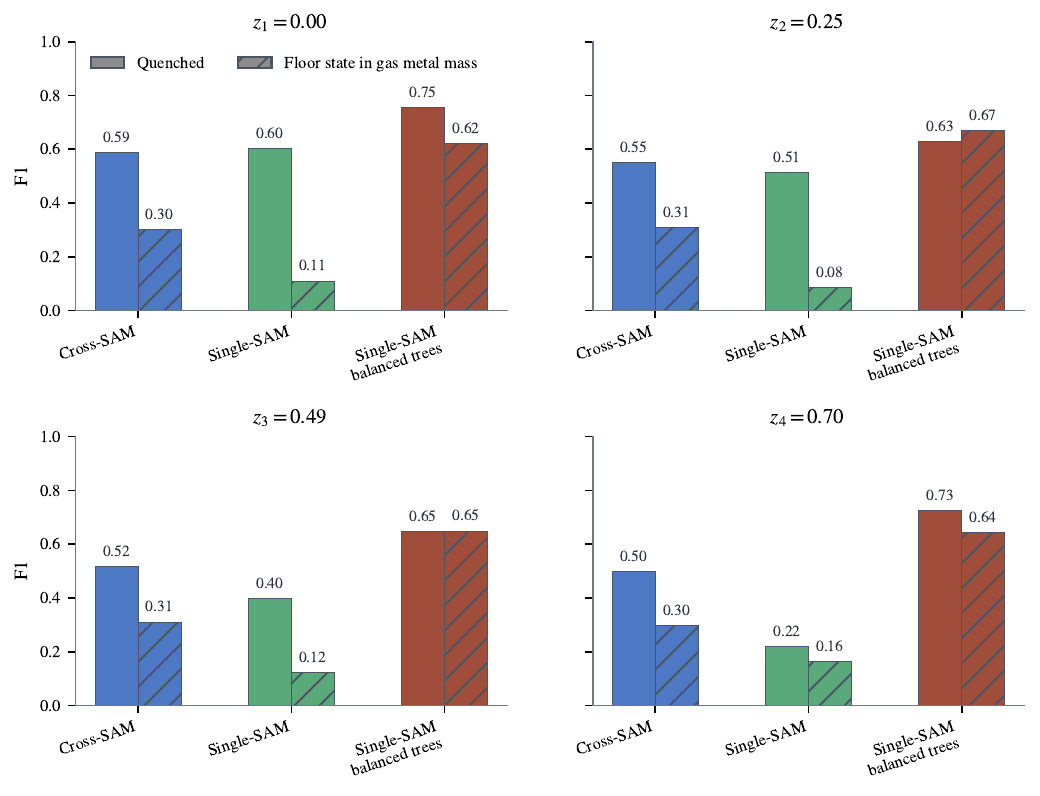}
  \caption{F1 scores for the quenched and floor states in gas metal mass under three data settings: the full cross SAM problem, a single SAM problem, and a single SAM balanced tree setting. Each panel corresponds to one of the first four redshift outputs. Within each panel, the two bar styles represent the two binary targets.}
  \label{fig:fig06}
\end{figure*}

\subsection{Decomposing Variation Across SAMs and Trees}

Figure~\ref{fig:sam_tree_variation_dumbbell} summarizes the central point of the controlled comparisons by separating two distinct sources of variation in the target properties. This reflects the intrinsic dispersion of these galaxy properties given the merger trees of the halos and SAM parameter sets. We refer to the variation measured across different merger trees at fixed \textsc{Galacticus} catalog as \emph{tree-to-tree} variation, and to the variation measured across different \textsc{Galacticus} SAM catalogs at fixed merger tree as \emph{catalog-to-catalog} variation. The former reflects differences in halo assembly history at fixed baryonic prescription, while the latter reflects differences induced by changing the SAM parameter vector at fixed merger history. The two half-violin distributions in each row therefore provide a direct visual comparison between tree-to-tree and catalog-to-catalog variation. The main conclusion is that the five targets are not limited by the same source of variation. Some are determined primarily by differences in halo assembly history, while others remain strongly sensitive to how the baryonic model responds to the same underlying merger history. This contrast is visible directly in the characteristic scatter ranges. Averaged over redshift, the model scatter is lowest for \(\Mstar\) and \(\Lz\) (\(\sim 0.24\)), larger for \(\Jang\) (\(\sim 0.41\)), and much larger for \(\Mzgas\) (\(\sim 0.53\)), while the corresponding \(\sSFR\) summary is complicated by the quenched branch. The value of the graph model is therefore not just that it improves prediction, but that it helps separate these regimes more clearly.

\(\Mstar\) and \(\Lz\) are influenced strongly by merger history, but they also remain sensitive to changes in the SAM parameters at fixed tree. In Figure~\ref{fig:sam_tree_variation_dumbbell}, both quantities show tree-to-tree and catalog-to-catalog spreads of comparable order, consistent with a regime in which assembly history sets much of the large scale trend while baryonic prescriptions still modulate the final outcome. They are therefore closely connected to halo growth, but not determined by it alone.

\(\Jang\) depends more strongly on differences between merger trees than on changes in the SAM parameters, which suggests that it retains more information about the merger history itself. Figure~\ref{fig:sam_tree_variation_dumbbell} shows this asymmetry directly: the tree-driven spread is broader than the catalog-driven spread, even though the model scatter remains at the level of \(\sim 0.41\). The modest improvement of the present model, therefore, likely reflects a limitation of the graph representation, which does not fully capture the dynamical details that matter for this quantity.

\(\Mzgas\) and \(\sSFR\) illustrate a different regime. Their difficulty is not well described as a single broad scatter around a smooth relation. For \(\Mzgas\), Figure~\ref{fig:sam_tree_variation_dumbbell} shows substantial spread even at fixed tree, consistent with strong sensitivity to cooling, feedback, and gas cycling prescriptions, and the model scatter remains much larger than for \(\Mstar\) or \(\Lz\). For \(\sSFR\), the star forming component varies across trees, but the quenched fraction changes substantially across catalogs. 

In order to further digest the results, we present Figure~\ref{fig:fig04} and Figure~\ref{fig:tree_state_variability}. Figure~\ref{fig:fig04} summarizes the catalog-level fractions of the discrete populations as a function of redshift, showing that the quenched fraction in \(\sSFR\) and the floor fraction in \(\Mzgas\) both vary substantially across catalogs.  At \(z=0\), the median quenched fraction is only 0.134 and the median floor fraction is 0.047, indicating a clearly imbalanced label distribution and setting up the motivation for the later single SAM balanced-tree control experiment. Figure~\ref{fig:tree_state_variability} then presents the corresponding tree-level view. Here, ``state'' refers to whether a galaxy is classified as quenched in \(\sSFR\) or as belonging to the floor population in \(\Mzgas\).The figure shows that many trees do not remain permanently in one state, but instead change state across catalogs; at \(z=0\), this occurs for 80.2\% of trees for the quenched label. These patterns suggest that the difficulty for \(\sSFR\) and \(\Mzgas\) arises not only from regression within a state, but also from identifying the correct physical state and its occupancy across the population.

\begin{table*}
\centering
\caption{{{
Performance summary for the mass-selected sample at $z=0$.
The model subscript denotes the training SAM/tree realization, while the entry following the arrow denotes the evaluation sample. The reference model is
$\mathrm{GNN}_{\mathrm{S04/T04}}\!\rightarrow\!\mathrm{S00/T00}$ as presented in bold format.
The Single-SAM models are trained on C1883/T04 and evaluated on the C1883/T00 sample; the tree-balanced variant uses equal numbers
of training trees with and without quenched galaxies.
For $\log(M_{Z,\mathrm{gas}})$ and $\log(\mathrm{sSFR})$, the final
column reports the $F_1$ score for identifying the floor population and quenched
systems, respectively. All non-ensemble rows report the performance of a single model.
}}}
\label{tab:z0_regression_summary}
\setlength{\tabcolsep}{5pt}
\begin{tabular}{llrrrrr}
\toprule
Property & Setting & bias & $\sigma$ & $\rho$ & $R^2$ & $F_1$\\
\midrule
$\log(M_{\star})$ & $\mathbf{GNN}_{\mathbf{S04/T04}}\!\rightarrow\!\mathbf{S00/T00}$ \textbf{ } & \textbf{0.009} & \textbf{0.225} & \textbf{0.978} & \textbf{0.957} & --\\
 & $\mathrm{MLP}^{z=0}_{\mathrm{S04/T04}}  \!\rightarrow\!  \mathrm{S00/T00}$ & 0.008 & 0.279 & 0.967 & 0.933 & --\\
 & $\mathrm{GNN}_{\mathrm{S04/T04}}\!\rightarrow\!\mathrm{S00/T04}$ (fixed tree) & 0.001 & 0.215 & 0.980 & 0.960 & --\\
 &  All-heteroscedastic-head $\mathrm{GNN}_{\mathrm{S04/T04}}  \!\rightarrow\!  \mathrm{S00/T00}$ & 0.017 & 0.224 & 0.978 & 0.957 & --\\
 & Single-SAM $\mathrm{GNN}_{\mathrm{C1883/T04}}\!\rightarrow\!\mathrm{C1883/T00}$ & -0.002 & 0.147 & 0.962 & 0.924 & --\\
 & Single-SAM $\mathrm{GNN}_{\mathrm{C1883/T04}}\!\rightarrow\!\mathrm{C1883/T00}$  (balanced trees) & -0.007 & 0.114 & 0.975 & 0.951 & --\\
 & 48-model prediction ensemble $\!\rightarrow\!\mathrm{S00/T00}$ & 0.005 & 0.218 & 0.979 & 0.959 & --\\
\addlinespace[2pt]

$\log(L_{z})$ & $\mathbf{GNN}_{\mathbf{S04/T04}}\!\rightarrow\!\mathbf{S00/T00}$ \textbf{ } & \textbf{0.009} & \textbf{0.243} & \textbf{0.971} & \textbf{0.943} & --\\
 & $\mathrm{MLP}^{z=0}_{\mathrm{S04/T04}}  \!\rightarrow\!  \mathrm{S00/T00}$ & 0.013 & 0.269 & 0.965 & 0.929 & --\\
 & $\mathrm{GNN}_{\mathrm{S04/T04}}\!\rightarrow\!\mathrm{S00/T04}$ (fixed tree) & -0.003 & 0.236 & 0.973 & 0.946 & --\\
 &  All-heteroscedastic-head $\mathrm{GNN}_{\mathrm{S04/T04}}  \!\rightarrow\!  \mathrm{S00/T00}$ & 0.021 & 0.243 & 0.971 & 0.942 & --\\
 & Single-SAM $\mathrm{GNN}_{\mathrm{C1883/T04}}\!\rightarrow\!\mathrm{C1883/T00}$ & -0.010 & 0.186 & 0.924 & 0.852 & --\\
 & Single-SAM $\mathrm{GNN}_{\mathrm{C1883/T04}}\!\rightarrow\!\mathrm{C1883/T00}$  (balanced trees) & -0.009 & 0.156 & 0.943 & 0.888 & --\\
 & 48-model prediction ensemble $\!\rightarrow\!\mathrm{S00/T00}$ & 0.007 & 0.237 & 0.972 & 0.945 & --\\
\addlinespace[2pt]

$\log(J)$ & $\mathbf{GNN}_{\mathbf{S04/T04}}\!\rightarrow\!\mathbf{S00/T00}$ \textbf{ } & \textbf{0.005} & \textbf{0.462} & \textbf{0.930} & \textbf{0.864} & --\\
 & $\mathrm{MLP}^{z=0}_{\mathrm{S04/T04}}  \!\rightarrow\!  \mathrm{S00/T00}$ & -0.000 & 0.462 & 0.929 & 0.860 & --\\
 & $\mathrm{GNN}_{\mathrm{S04/T04}}\!\rightarrow\!\mathrm{S00/T04}$ (fixed tree) & 0.006 & 0.424 & 0.940 & 0.883 & --\\
 &  All-heteroscedastic-head $\mathrm{GNN}_{\mathrm{S04/T04}}  \!\rightarrow\!  \mathrm{S00/T00}$ & 0.021 & 0.462 & 0.929 & 0.863 & --\\
 & Single-SAM $\mathrm{GNN}_{\mathrm{C1883/T04}}\!\rightarrow\!\mathrm{C1883/T00}$ & -0.014 & 0.419 & 0.892 & 0.795 & --\\
 & Single-SAM $\mathrm{GNN}_{\mathrm{C1883/T04}}\!\rightarrow\!\mathrm{C1883/T00}$  (balanced trees) & -0.017 & 0.412 & 0.892 & 0.794 & --\\
 & 48-model prediction ensemble $\!\rightarrow\!\mathrm{S00/T00}$ & -0.000 & 0.459 & 0.930 & 0.866 & --\\
\addlinespace[2pt]

$\log(M_{Z,\mathrm{gas}})$ & $\mathbf{GNN}_{\mathbf{S04/T04}}\!\rightarrow\!\mathbf{S00/T00}$ \textbf{ } & \textbf{0.176} & \textbf{0.694} & \textbf{0.905} & \textbf{0.806} & \textbf{0.303}\\
 & $\mathrm{MLP}^{z=0}_{\mathrm{S04/T04}}  \!\rightarrow\!  \mathrm{S00/T00}$ & -0.052 & 1.233 & 0.743 & 0.489 & 0.000\\
 & $\mathrm{GNN}_{\mathrm{S04/T04}}\!\rightarrow\!\mathrm{S00/T04}$ (fixed tree) & 0.197 & 0.746 & 0.898 & 0.791 & 0.372\\
 &  All-heteroscedastic-head $\mathrm{GNN}_{\mathrm{S04/T04}}  \!\rightarrow\!  \mathrm{S00/T00}$ & 0.060 & 0.771 & 0.895 & 0.799 & 0.000\\
 & Single-SAM $\mathrm{GNN}_{\mathrm{C1883/T04}}\!\rightarrow\!\mathrm{C1883/T00}$ & -0.031 & 0.908 & 0.399 & 0.150 & 0.235\\
 & Single-SAM $\mathrm{GNN}_{\mathrm{C1883/T04}}\!\rightarrow\!\mathrm{C1883/T00}$  (balanced trees) & -0.005 & 0.694 & 0.567 & 0.319 & 0.621\\
 & 48-model prediction ensemble $\!\rightarrow\!\mathrm{S00/T00}$ & 0.170 & 0.689 & 0.907 & 0.811 & 0.310\\
\addlinespace[2pt]

$\log(\mathrm{sSFR})$ & $\mathbf{GNN}_{\mathbf{S04/T04}}\!\rightarrow\!\mathbf{S00/T00}$ \textbf{ } & \textbf{0.042} & \textbf{0.221} & \textbf{0.735} & \textbf{0.522} & \textbf{0.588}\\
 & $\mathrm{MLP}^{z=0}_{\mathrm{S04/T04}}  \!\rightarrow\!  \mathrm{S00/T00}$ & -0.062 & 0.335 & 0.510 & 0.115 & 0.042\\
 & $\mathrm{GNN}_{\mathrm{S04/T04}}\!\rightarrow\!\mathrm{S00/T04}$ (fixed tree) & 0.043 & 0.229 & 0.732 & 0.519 & 0.575\\
 &  All-heteroscedastic-head $\mathrm{GNN}_{\mathrm{S04/T04}}  \!\rightarrow\!  \mathrm{S00/T00}$ & 0.008 & 0.236 & 0.760 & 0.576 & 0.023\\
 & Single-SAM $\mathrm{GNN}_{\mathrm{C1883/T04}}\!\rightarrow\!\mathrm{C1883/T00}$ & 0.062 & 0.377 & 0.475 & 0.200 & 0.575\\
 & Single-SAM $\mathrm{GNN}_{\mathrm{C1883/T04}}\!\rightarrow\!\mathrm{C1883/T00}$  (balanced trees) &0.023 & 0.250 & 0.693 & 0.472 & 0.754\\
 & 48-model prediction ensemble $\!\rightarrow\!\mathrm{S00/T00}$ & 0.045 & 0.221 & 0.732 & 0.516 & 0.586\\
\addlinespace[2pt]

\bottomrule
\end{tabular}
\end{table*}

\subsection{Head Design and Control Experiments}
{{Table~\ref{tab:z0_regression_summary} compares the fixed reference realization with several complementary settings at $z=0$. To distinguish the data used to train a model from those used to evaluate it, we extend the notation introduced in Section~\ref{sec:data_split} by using the model subscript only for the training SAM/tree pair and placing the evaluation pair to the right of an arrow. The default reference evaluation is therefore written as $\mathrm{GNN}_{\mathrm{S04/T04}}\!\rightarrow\!\mathrm{S00/T00}$. The fixed-tree evaluation is written as $\mathrm{GNN}_{\mathrm{S04/T04}}\!\rightarrow\!\mathrm{S00/T04}$: it uses the same trained checkpoint and the same 600 held-out SAM catalogs, and changes only the evaluation chunk from T00 to the training chunk T04.}}

{{The remaining rows address three distinct types of comparison. The MLP baseline and the all-heteroscedastic-head GNN modify the model input or output architecture. The fixed-tree, single-SAM, and quenched-tree-balanced settings modify the evaluation or training sample. The ensemble forms a separate category because it averages predictions from models trained on 48 different realizations before computing the evaluation metrics. The MLP baseline, $\mathrm{MLP}^{z=0}_{\mathrm{S04/T04}}\!\rightarrow\!\mathrm{S00/T00}$,  uses the halo properties at $z=0$ together with the SAM parameter vector, without the merger tree graph. The all-heteroscedastic-head GNN retains the graph-based model but replaces the MoE heads used for $\Mzgas$ and $\sSFR$ with heteroscedastic regression heads. The single-SAM controls use SAM catalog C1883 \footnote{{{We randomly pick a SAM catalog from S04 and denote it as CXX, where XX is our SAM model ID.}}} alone: both are trained on T04 and evaluated on the disjoint T00 tree chunk. For the single-SAM  (balanced trees) setting, in which the training sample is further resampled so that the \(\sSFR\) labels are balanced at a 1:1 ratio between quenched and non-quenched cases, i.e., we artificially downsample to change the fraction of them such that they have comparable number of galaxies in the training, following standard practice for imbalanced classification problems \citep{he2009learning}.}}

For \(\Mstar\), \(\Lz\), and \(\Jang\), {{the all-heteroscedastic-head GNN}} is already sufficient, consistent with the fact that these targets are described reasonably well by a single smooth conditional distribution. The clearest head dependence appears instead for \(\Mzgas\) and \(\sSFR\), where the problem involves state identification as well as continuous regression. For \(\Mzgas\), the all-heteroscedastic model has essentially no ability to identify the floor population (\(F_1=0.000\)), whereas the {{reference}} model reaches $F_1=0.303$. For \(\sSFR\), the heteroscedastic model can still fit the continuous distribution reasonably well, but it largely fails to identify quenched systems (\(F_1=0.023\) versus \(0.588\) for the reference model). For these targets, evaluation should therefore account for the discrete state structure of the distribution rather than relying on regression metrics alone. 

Figure~\ref{fig:fig06} then shows which part of this difficulty can be reduced by changing the data setting. A single-SAM setting does not improve the classification scores much, whereas balancing the tree sample does. This indicates that the limiting factor is not only variation in state occupancy across catalogs, but also the uneven distribution of trees and labels in the sample itself.

\subsection{Limitations and future {work}}
Although our model has shown the capability of emulating SAMs, we note that there are several limitations in our model and results. 

First, the present model is trained within one simulation and one SAM framework (Galacticus), so the conclusions concern generalizations inside that setup rather than transfer across cosmologies, halo finders, or semi-analytic formalisms. 

Second, the graph uses a relatively simple representation of the merger tree and does not include explicit edge attributes, such as orbital properties, nor does it use separate message passing rules for progenitor-descendant and host-satellite relations. That simplification is likely most restrictive for \(\Jang\), which appears to depend more strongly on dynamical information that is only partially encoded by the present tree representation. 

Third, the remaining classification performance for \(\sSFR\) and \(\Mzgas\) shows that the state classification is still not captured adequately in the hardest cases. 

Fourth, the current pipeline does not learn autonomously which nodes should host target galaxy predictions. Instead, predictions are made only for the subset of halo nodes associated with galaxies, so occupancy is treated as known rather than inferred by the model. A natural extension would be to model halo occupancy jointly with the galaxy properties, for example, with a generative formulation that first samples whether a halo hosts a galaxy and then generates its properties conditional on the tree and the SAM parameters. Related generative approaches could also represent the full conditional distribution over occupied halos and galaxy states more flexibly than the current deterministic masking step. 

In addition, the current graph construction does not yet exploit the full merger history available in the trees. The subsampling may yield unexpected noise in the final output. {Note that} our model only utilizes the internal halo properties. However, we know that the external environment also plays a crucial role in shaping the galaxy properties \citep{dressler1980galaxy, kauffmann2004environment, peng2012quenching, Wechsler_Tinker_2018}. {{The current \textsc{Galacticus} model does not explicitly incorporate any environment parameter. But part of this information can enter the model through the merger tree structure of the dark matter halos and further propagate to our GNN construction. Adding additional suitable environmental descriptors could improve both the \textsc{Galacticus} modeling in terms of predicting galaxy properties, and the surrogate's fidelity to reproduce \textsc{Galacticus} output, but this is beyond the scope of the current work.}}

These limitations suggest a clear direction for future work. Better surrogates will likely require broader training coverage, more explicit treatment of state structure in the outputs, explicit learning of halo occupancy rather than relying on a fixed mask, and graph representations that retain more of the information carried by mergers. In practice, the main goal is to make these models more reliable for repeated SAM evaluation across large samples and broad parameter spaces.

\section{Conclusion}\label{sec:conclusion}
We have shown that a graph neural network can serve as an effective surrogate for \textsc{Galacticus} predictions of multiple galaxy properties across multiple redshift outputs, conditioned on halo merger trees and SAM parameters. In the fiducial Multi-SAM GNN setting, the model recovers several galaxy properties with good accuracy over a wide redshift range, with the most stable performance obtained for stellar mass and luminosity. The results also show that the prediction difficulty depends strongly on the target quantity. Properties with comparatively smooth conditional structure, such as $\Mstar$ and $\Lz$, are recovered more robustly, whereas targets with distinct branches or floor-dominated structure, such as $\sSFR$ and $\Mzgas$, benefit substantially from the state-aware output design. Ensemble averaging provides an additional but modest improvement in the continuous regression metrics, while leaving the discrete state identification performance broadly unchanged.

These results indicate that surrogate models informed by merger tree structure can provide a practical approximation to repeated SAM evaluation in applications that require many forward model calls, including parameter exploration, calibration, and mock catalog generation. {In that regime, the surrogate is expected to be at least one order of magnitude faster than rerunning the full SAM on the same merger tree input. For instance, processing the whole UchuuMicro using the Galacticus model on a single CPU can take roughly 10 hours, while our GNN model needs less than 1 minute. On the other hand, we should note that this comparison may not be fully fair since the hardware infrastructure and the model output of galaxy properties are different, but our model demonstrates that for a large scale problem of multiple trees in high dimensional parameter space, it is still a feasible choice {to develop} GNN-based surrogate model.} At the same time, the present model should be viewed as a surrogate within the current simulation and SAM setup rather than as a replacement for full semi-analytic modeling in all regimes. Its performance remains strongest for smoothly varying targets and weaker in the rare object and state classification regimes.

Natural next steps include extending the method to a broader range of simulation and SAM settings, improving the treatment of discrete galaxy states and halo occupancy, and adopting richer graph representations that retain more of the information carried by merger histories and environments.

\section{Acknowledgements}
This work is supported by the National Key R\&D Program of China (2023YFA1607800, 2023YFA1607804, 2023YFA1605600), the National Science Foundation of China (No. 12373003), “the Fundamental Research Funds for the Central Universities”, 111 project No. B20019, Shanghai Natural Science Foundation, grant No.19ZR1466800, and by Office of Science and Technology, Shanghai Municipal Government (grant Nos. 24DX1400100, ZJ2023-ZD-001).
We acknowledge the science research grants from the China Manned Space Project with Nos. CMS-CSST-2021-A02 \& CMS-CSST-2025-A04, and Yangyang Development Fund.

\section*{Data Availability}
{{The \textsc{SAM2Galaxy GNN} inference package, including the released model
checkpoints, example inputs and outputs, and the exact SAM catalog and merger
tree split manifests used in this work, is archived on Zenodo
\citep{li2026sam2galaxy}. The halo merger tree input is available as a release
asset linked from the archived package documentation.}}

\bibliographystyle{aasjournalv7}
\bibliography{main}
\appendix

\section{{{Variation across single-model realizations}}}\label{app:realization_variation}

\begin{figure*}
  \centering
  \includegraphics[width=0.92\textwidth]{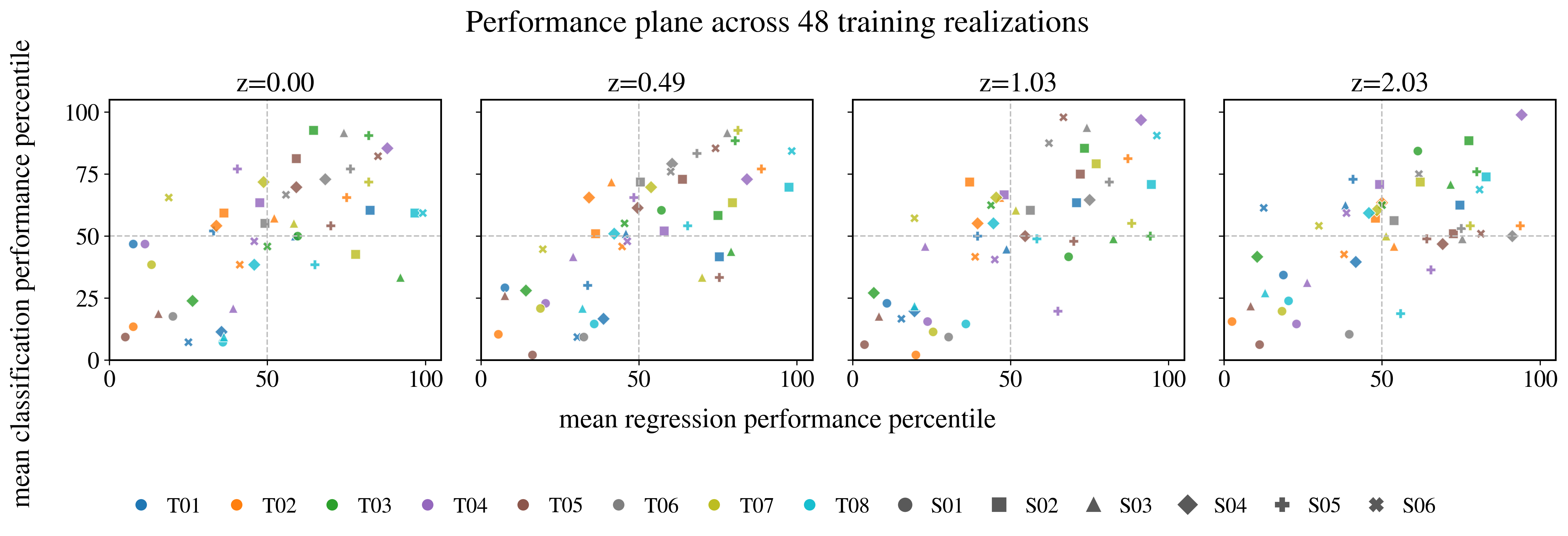}
  \caption{{{Regression--classification performance plane for the 48 individual SAM-group/tree-chunk realizations at four representative redshifts. Colors identify tree chunks and symbols identify SAM groups. Both coordinates are percentile ranks normalized to 0--100 within the 48 realizations, with 100 denoting the best relative performance. These coordinates show relative ordering rather than the absolute size of the differences: at $z=0$, for example, the stellar mass scatter varies only from $0.223$ to $0.231$, indicating that there is no substantial variations across these individual realizations.}}}
  \label{fig:performance_plane}
\end{figure*}

{{Figure~\ref{fig:performance_plane} places all 48 individual models in a regression--classification performance plane. For each redshift, the horizontal coordinate is the mean performance percentile of the continuous scatter across the five targets, and the vertical coordinate is the mean performance percentile of the $F_1$ scores of the quenched and floor states in gas metal mass; both axes are oriented so that larger values indicate better performance. Colors distinguish the eight tree chunks, symbols distinguish the six SAM groups. }}

\section{{{Stellar mass residual distributions in individual SAM catalogs}}}
\label{app:halo_mass_catalogs}

\begin{figure*}
  \centering
  \includegraphics[width=0.98\textwidth,keepaspectratio]{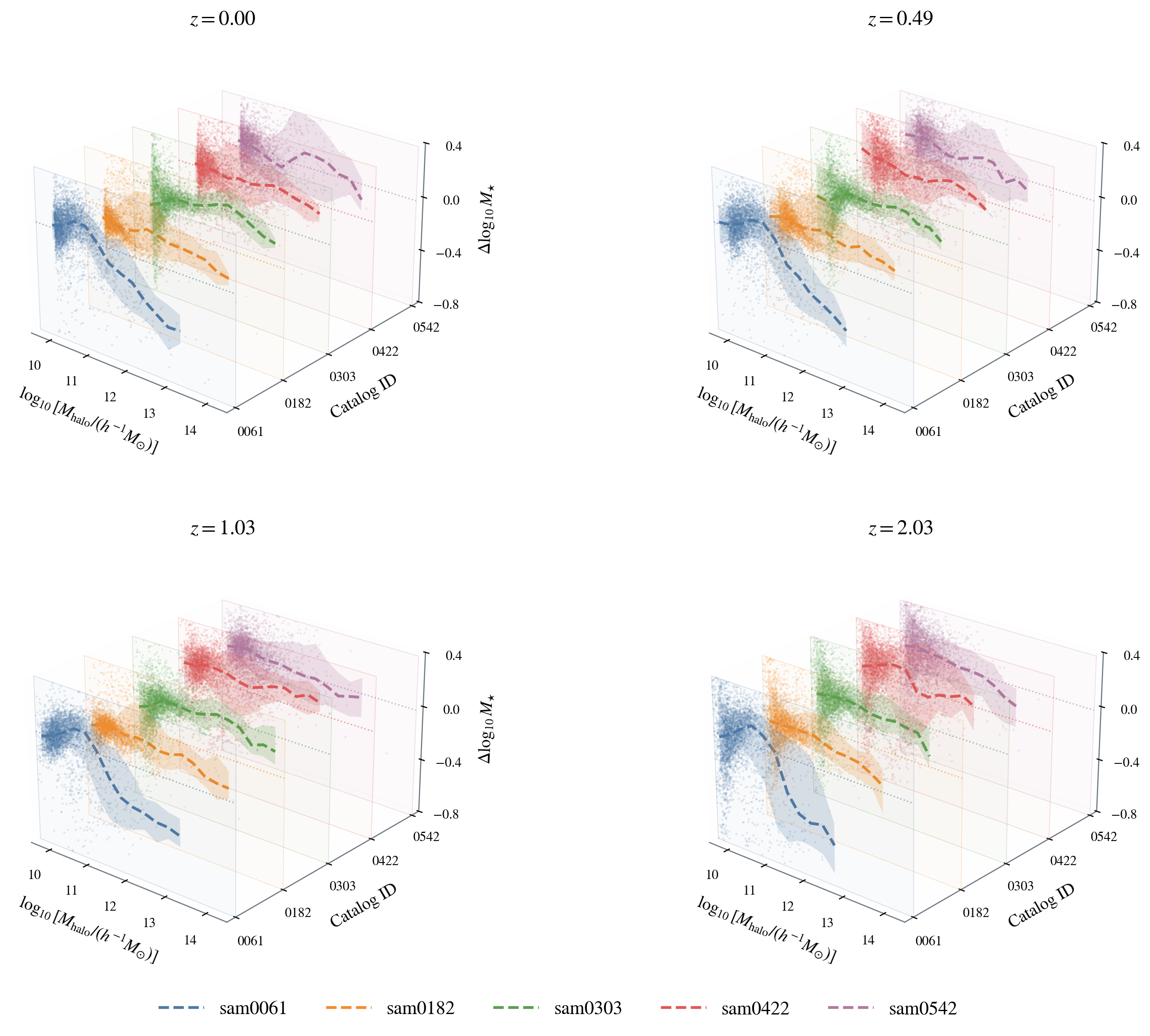}
  \caption{{{Residuals of the stellar mass prediction for individual galaxies as a function of halo mass for five test SAM catalogs at four redshifts. Dashed curves and shaded regions show the median and 16th--84th percentile range within each halo mass bin. For visual clarity, the displayed points are randomly subsampled, while the summary curves use all available galaxies.}}}
  \label{fig:mstar_residual_5catalog_stack}
\end{figure*}

{{The halo mass diagnostic in the main text summarizes the 600 test SAM catalogs
with statistics calculated for each catalog. To expose the underlying
distributions for individual galaxies,
Figure~\ref{fig:mstar_residual_5catalog_stack} shows five illustrative test
catalogs. The negative shift at large halo mass is visible in each of the five catalogs,
although its onset and dispersion vary with the SAM realization and redshift.
This view of individual catalogs confirms that the trend in
Figure~\ref{fig:mstar_residual_halo_mass}.}}

\section{Additional Prediction Panels}\label{Appendix:A}

\begin{figure*}
  \centering
  \includegraphics[width=0.9\textwidth]{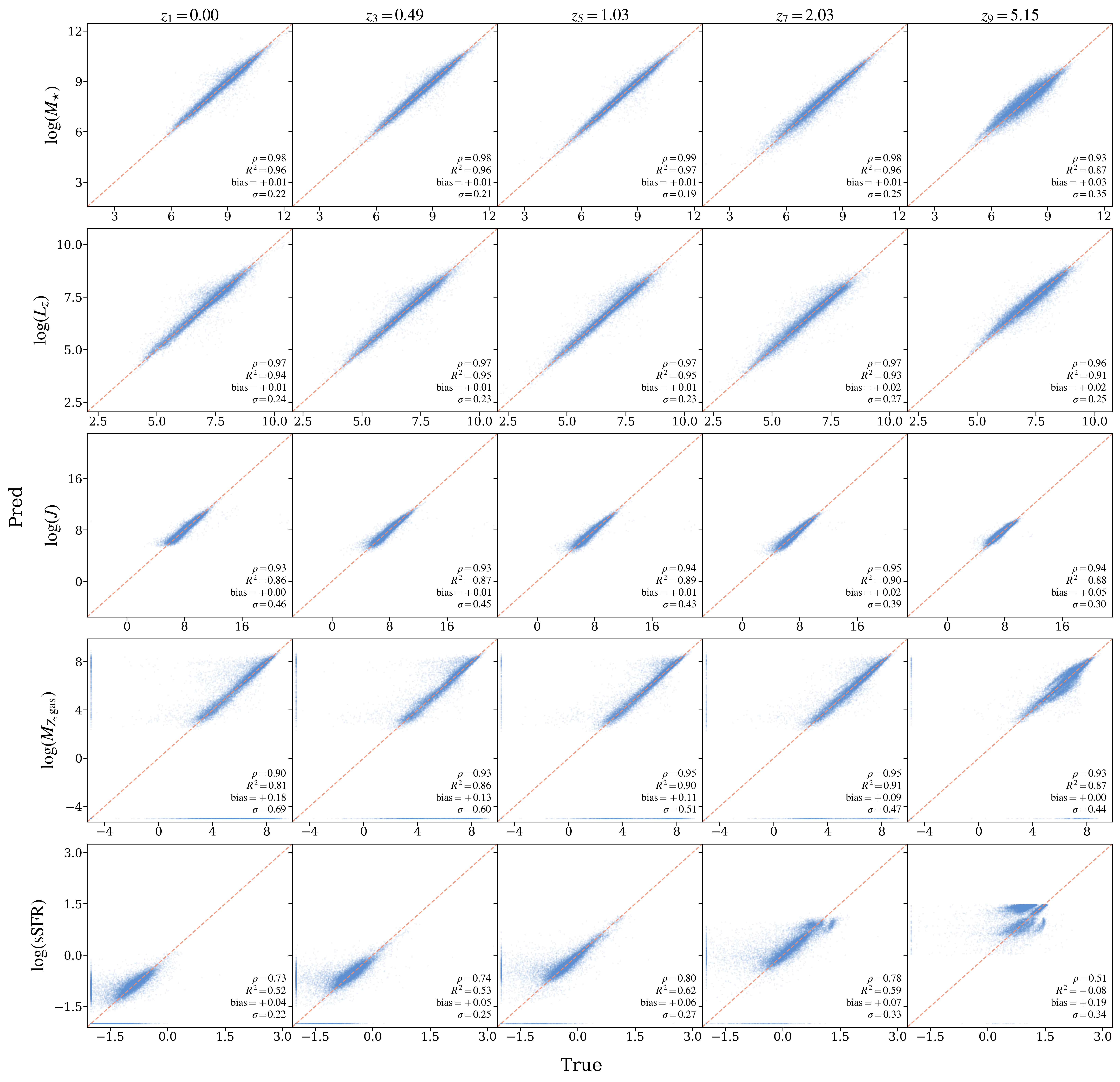}
  \caption{Predicted versus true values for the five target properties at five representative redshift outputs. Columns correspond to different redshifts and rows correspond to different target properties. Each panel shows the one-to-one relation together with the corresponding predicted and true values from the fiducial Multi-SAM GNN.}
  \label{fig:scatter_grid_overview}
\end{figure*}

Figure~\ref{fig:scatter_grid_overview} summarizes the pointwise prediction results for all five target properties at five representative redshift outputs. By displaying the predicted versus true relations in a single grid, it provides a compact overview of how the prediction quality changes across both target quantity and redshift. The figure is consistent with the main results in Section~\ref{sec:results}: stellar mass and luminosity remain comparatively tight over most of the sampled redshift range, whereas angular momentum shows broader scatter and $\sSFR$ and $M_{Z,\mathrm{gas}}$ retain more structured behavior.

\end{document}